\documentclass[a4paper,11pt]{article}
\usepackage{pos}
\usepackage{graphicx} % Only load graphicx; 'graphics' is not needed
\usepackage{tikz}
\usetikzlibrary{3d,calc,arrows.meta,decorations.markings}
\usepackage{tikz-feynman}
\tikzfeynmanset{compat=1.1.0}
\usepackage{feynmp}
\usepackage{feynmp-auto}
\usepackage{caption}
\usepackage{subcaption}
\usepackage{multirow}

\title{A comprehensive study of $B$, $B_s$ \& $B_c$ meson semitauonic modes in potential quark model}
\author*[a]{Sonali Patnaik}

\affiliation[a]{Indian Institute of Technology, North Guwahati, Guwahati 781039, Assam, India}

\emailAdd{spatnaik@iitg.ac.in}

\abstract{In this work, we derive the form factors and compute the branching fractions for the semitauonic decay modes,
$B \to D^{(*)}\,\tau\,\nu_{\tau}$, $B_s \to D_s^{(*)}\,\tau\,\nu_{\tau}$, $B_c \to \eta_c\,(J/\psi)\,\tau\,\nu_{\tau}$, and 
$B_c \to D^{(*)}\,\tau\,\nu_{\tau}$ within the \emph{Relativistic Independent Quark (RIQ) Model}, 
emphasizing a quark potential model based analysis of these transitions. We outline the essential elements of the model, incorporating corrections from residual interactions and 
center-of-mass motion, and perform a comprehensive study of the form factors across the full physical kinematic range of $q^2$. The resulting predictions demonstrate consistency and good agreement with existing theoretical approaches and experimental measurements. Motivated by recent observations of polarization observables at LHCb and Belle, we further evaluate these quantities within our framework and find results compatible with Standard Model (SM) expectations. The predictions presented here serve as theoretical input in decay channels for which lattice QCD results remain limited, offering guidance for future experimental and Lattice efforts. Thus, semileptonic $B$ decays continue to serve as precise 
and discerning probes of the fundamental mechanisms governing flavor transitions in the SM.}

\FullConference{Proceedings of the Corfu Summer Institute 2025 "School and Workshops on Elementary Particle Physics and Gravity" (CORFU2025)\\
27 April - 28 September, 2025\\
Corfu, Greece\\}

\begin{document}
\maketitle
Weak decays of $B$, $B_s$, and $B_c$ mesons containing a bottom ($b$) quark and spectator quarks ($ u,d,s,c$) serve as excellent probes of understanding physics within the SM and beyond (BSM). These decays test lepton flavor universality, as their SM predictions arise from clean tree-level processes. Recent deviations observed in $b \to c\ell\nu$ and $b \to s\ell\nu$ transitions have generated interest in flavor anomalies. Measurements of $\tau$ polarization $P_{\tau}(D^*)$~\cite{Belle:2016dyj,Belle:2019ewo} and fraction of longitudinal polarization $F_L^{D^*}$ are consistent with SM expectations~\cite{Huang:2018nnq,Bhattacharya:2018kig}, as reaffirmed by LHCb~\cite{LHCb:2023ssl}. Since $B_s \to D_s^{(*)}\ell\nu_\ell$ and $B \to D^{(*)}\ell\nu_\ell$ proceed via the same $b \to c \ell \nu_\ell$ transition, SU(3) flavor symmetry~\cite{Khlopov:1978id}implies similar semileptonic behavior. Any new physics effects in $B \to D^{(*)}\ell\nu$ should thus also appear in $B_s \to D_s^{(*)}\ell\nu$ channels due to SU(3) flavor symmetry effects. Observables such as $P_\tau(D_s^{(*)})$ and $F_L(D_s^*)$ therefore provide complementary sensitivity to BSM effects, while the sizable decay width of $B_c \to \eta_c\, (J/\psi)\, \bar{\ell}\,\nu_\ell$ makes it another promising probe~\cite{Chang:1992pt,Dey:2025xdx} . On the theoretical side, weak decays offer a crucial means to determine transition form factors using non-perturbative methods such as the Bethe–Salpeter equation (BSE), QCD sum rules (QCDSRs), and Lattice QCD (LQCD). Combining HQET and QCDSRs allows estimates for $B \to D \tau \nu_{\tau}$, though with limited precision~\cite{Bernlochner:2021vlv}. Besides, QCD-inspired quark models remain valuable tools when they successfully reproduce experimental data across hadron sectors. Motivated by ongoing $B$-physics efforts, we therefore investigate exclusive semileptonic decays of $b$-flavored mesons using the Relativistic Independent Quark (RIQ) model. We compute the semileptonic decays: $B \to D^{(*)}$, $B_s \to D_s^{(*)}$,  $B_c \to \eta_c\,(J/\psi)$ and $B_c \to D^{(*)}$ focusing on $\tau$ modes governed by six Lorentz invariant form factors. These are extracted in the RIQ model as overlap integrals of meson wave functions~\cite{Nayak:2021djn,Nayak:2024esq}. We evaluate them over the full physical kinematic range of $q^2$ and present branching fractions, form factors and angular obseravbles for $\tau$ modes. In the absence of sufficient LQCD data, our analysis offers valuable SM predictions that enhance the understanding of semileptonic $B$ decays, particularly in channels involving third-generation leptons.
\section{RIQ model framework}
\label{sec:RIQ}
Studying exclusive semileptonic decays involving non-perturbative hadronic matrix elements can be defined using confining interaction potential formulation. In this context, we provide a concise theoretical predictions adopting the RIQ model (RIQM) framework. The RIQM is grounded in a confining harmonic potential in the equally mixed scalar-vector form~\cite{Patnaik:2017cbl,Nayak:2021djn,Patnaik:2023efe},
\begin{equation}
U(r)=\frac{1}{2}\left(1+\gamma^0\right)\,V(r)\,, \,\,\, V(r)=(ar^2+V_0).
\label{eq:harmonicpotential}
\end{equation}
Here, $r$ represents a state dependent length parameter, $\gamma^0$ denotes the time-like Hermitian matrix, $a$ and $V_0$ are potential parameters. These parameters are fixed during the hadron spectroscopy~\cite{Patnaik:2025fry}. This confining interaction is believed to provide phenomenologically the zeroth order quark dynamics inside the hadron-core through the quark Lagrangian density,
This confining interaction is believed to provide phenomenologically the zeroth order quark dynamics inside the hadron-core through the quark Lagrangian density, 
\begin{equation}
{\cal L}^{0}_{q}(x)={\bar \psi}_{q}(x)\;[\;{\frac{i}{2}}\gamma ^{\mu}
\partial _{\mu}-m_{q}-U(r)\;]\;\psi _{q}(x)
\label{ld},
\end{equation}
leading to the Dirac equation for individual quark \& antiquark systems. Since first principle QCD calculations remain challenging due to its non-perturbative nature therefore RIQM derives observable hadron properties from constituent quark dynamics. The model  characterizes quark confinement through the chosen potential in Eq.~\eqref{eq:harmonicpotential}, representing multi-gluon interactions with a specified Lorentz structure. Residual effects such as one-gluon exchange (OGE) at short distances and quark–pion coupling in the light sector are treated perturbatively~\cite{Barik:1986mq,Barik:1987zb}. As the present work focuses on heavy hadrons pionic contribution becomes negligible and only the OGE term is considered. Therefore in the next section we briefly describe the OGE within the RIQM.  
\subsection{One gluon-exchange correction}
\label{subsec:oge}
% First diagram (gluon exchange)
\begin{figure}
    \centering
    \begin{tikzpicture}
        \begin{feynman}
        % Define the four endpoints of the fermion lines
        \vertex (a1) at (-1.5,1);
        \vertex (a2) at (1.5,1);
        \vertex (b1) at (-1.5,-1);
        \vertex (b2) at (1.5,-1);
        
        % Define new intermediate vertices along the fermion lines
        \vertex (m1) at (-0,1);  % Middle of upper fermion line
        \vertex (m2) at (-0,-1); % Middle of lower fermion line

        % Draw the Feynman diagram
        \diagram* {
            (a1) -- [plain] (m1) -- [plain] (a2), % Upper fermion line
            (b1) -- [plain] (m2) -- [plain] (b2), % Lower fermion line
            (m1) -- [gluon] (m2), % Gluon connecting middle points
        };
        \end{feynman}
    \end{tikzpicture}
    \hspace{1.5cm}
    % Second diagram (self-energy correction)
    \begin{tikzpicture}
        \begin{feynman}
            \vertex (a) at (-1.1,0);
            \vertex (b) at (1.1,0);
            \vertex (c) at (0.7,0);
            \vertex (d) at (-0.7,0);
            \vertex (f) at (-1.1,-1.0);
             \vertex (g) at (1.1,-1.0);

             \diagram* {
                (a) -- [plain] (c),
                (d) -- [plain] (b),
                (c) -- [plain] (d),
                (c) -- [gluon, bend right=80] (d),
                (f) -- [plain] (g),
            };
        \end{feynman}
    \end{tikzpicture}
    \caption{One gluon exchange contribution to the energy of $q\bar{q}$ configuration}
    \label{fig:oge}
\end{figure}
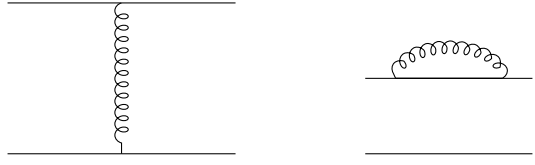
Within the meson core quarks experience the effective potential $U(r)$ along with a residual weak OGE interaction as depicted in Fig.~\ref{fig:oge} from the interaction Lagrangian density,
\begin{equation}
    {\cal L}_I^g=\sum_a J_i^{\mu a}(x)A_\mu^a(x),
\end{equation}
where $A_{\mu}^a(x)$ are vector-gluon fields and $J^{\mu a}(x)$ is the ith-quark colour-current. Since at small distances the quarks should be almost free, it is reasonable to calculate thw shift in energy of the meson-core arising out of the quark interaction energy due to its coupling to the coloured gluons, using a first order perturbation theory. Such an approach leads to the colour-electric and colour-magnetic energy shifts~\cite{Barik:1986mq,Barik:1987zb}.
\begin{align}
    (\Delta E_M)_g &= (\Delta E_M)_g^{\mathcal{E}} + (\Delta E_M)_g^{\mathcal{M}}, \label{eq:total_oge}\\[6pt]
    (\Delta E_M)_g^{\mathcal{E}} &= \alpha_s \sum_{i,j} 
    \bigg\langle \sum_a \lambda_i^a \lambda_j^a \bigg\rangle 
    \frac{1}{\sqrt{\pi} R_{ij}}
    \left( 1 - \frac{\alpha_i + \alpha_j}{R_{ij}^2} 
    + \frac{3 \alpha_i \alpha_j}{R_{ij}^4} \right), \label{eq:oge_electric}\\[6pt]
    (\Delta E_M)_g^{\mathcal{M}} &= \alpha_s \sum_{i<j} 
    \bigg\langle \sum_a \lambda_i^a \lambda_j^a \, \boldsymbol{\sigma}_i \cdot \boldsymbol{\sigma}_j \bigg\rangle 
    \frac{256}{9\sqrt{\pi}} 
    \frac{1}{(3E'_i + m'_i)(3E'_j + m'_j)} 
    \frac{1}{R_{ij}^3}. \label{eq:oge_magnetic}
\end{align}
Here $\lambda_i^a$ denote the Gell-Mann SU(3) matrices and $\alpha_s$ represents the effective scale-dependent strong coupling constant. Considering the specific flavor and spin configurations of ground-state mesons, the corresponding OGE energy correction $(\Delta E_M)_g$ is obtained. This would give the total energy of $(q_i\bar{q}_j)$ system in its ground state as,
\begin{equation}
    E_M=E_M^0+[(\Delta E_M)_g^{\cal E}+(\Delta E_M)_g^{\cal M}],
\end{equation}
where $E_M^0$ is the zeroth order energy for a ground-state meson $M$ arising out of the binding energies of constituent quark and antiquark confined independently by average potential $U(r)$. We further incorporate the correction from the spurious center-of-mass motion of $q \bar q$ system, detailed below.
\subsection{Centre of mass momentum and meson mass}
In this shell-type relativistic independent quark model, the motion of individual quarks within the hadron core does not inherently result in a state with a well-defined total momentum as required for a physically consistent hadron state. Therefore, the energy contribution from the spurious center-of-mass motion must be accounted as an additional correction to the hadron energy, obtained from the individual quark binding energy over and above the perturbative corrections discussed in~\ref{subsec:oge}. This correction has been thoroughly detailed in the earlier work~\cite{Barik:1986mq}.

In such an approach, the static meson-core state with core-centre at X is decomposed into components $\Phi(P)$ of plane-wave momentum eigen states as,
\begin{equation}
|M(X)\rangle_c=\int\frac{d^3P}{W_M(P)}exp(iP.X)\Phi_M(P)|M(P)\rangle.
\end{equation}
Taking its inverse relation along with the normalisation, one can obtain the momentum profile function $\Phi_M(P)$ as,
\begin{equation}
\Phi_M^2(P)=\frac{W_M(P)}{(2\pi)^3}\tilde{I}_M(P),
\end{equation}
where $\tilde{I}_M(P)$ is the fourier-transform of the Hill-wheeler overlap function~\cite{Wong:1980ce} given by,
\begin{eqnarray}
\tilde{I}_M(P)=&&\Bigg(\frac{r_{0q}^2}{2\pi}\Bigg)^{3/2}exp(-P^2r_{0q}^2/2)(1-6C_q+15C_q^2)\nonumber\\
&&\Bigg[1+\frac{P^2r_{0q}^2(2C_q-10C_q^2+C_q^2P^2r_{0q}^2)}{(1-6C_q+15C_q^2)}\Bigg],
\end{eqnarray}
when, $C_q=(E_q'-m_q')/6(3E_q'+m_q')$. We now estimate the center-of-mass momentum $P$ as,
\begin{eqnarray}
    \langle P^2\rangle=\int d^3P\tilde{I}_M(P)P^2
    =\sum_q\langle p^2 \rangle.
    \label{eq:comm1}
\end{eqnarray}
$\langle p^2 \rangle$ is the average value of the square of the individual quark-momentum taken over $1S_{1/2}$ single-quark state and is given by,
\begin{equation}
    \langle p^2 \rangle_q=\frac{(11E'_q+m'_q)(E_q^{2'}-m_q^{2'})}{6(3E'_q+m'_q)}.
    \label{eq:comm2}
\end{equation}
Finally taking into account the centre-of-mass motion of $(q_i\bar{q_j})$ system with the center-of-mass momentum $P$ as in Eq.~\ref{eq:comm1} \&~\ref{eq:comm2}, one can obtain the physical mass of $(q_i\bar{q_j})$ meson in its ground state as,
\begin{equation}
    m_M=[E^2_M-\langle P^2\rangle_M]^{1/2}.
    \label{eq:mmass}
\end{equation}
By including residual interaction effects and center-of-mass corrections, the model reliably reproduces the meson spectrum across various S-wave states through fitted quark masses and potential parameters, in good agreement with experimental data~\cite{Nayak:2025amk,Dash:2023ohf}. The extended RIQM framework thus provides a robust phenomenological tool for studying hadronic properties. Since the focus of this work is on semileptonic decay observables, we abstarin from presenting a detailed mass spectrum analysis and instead concentrate on predicting branching fractions and form factors for the relevant decay channels.
\section{Theoretical setup for the weak decays in RIQM}
\label{subsec:TF}
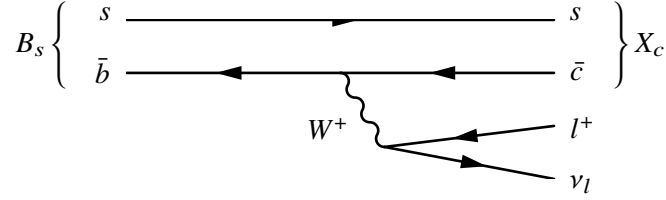
\begin{figure}[h!]
 \centering
 \begin{fmffile}{inclusive}
  \begin{fmfgraph*}(160, 60)
    \fmfset{arrow_len}{10}
    \fmfstraight
    \fmfleft{i4,i3,i2,i1}
    \fmfright{o4,o3,o2,o1}
    % fermions
    \fmffreeze
    \fmf{fermion}{i1,o1}
    \fmf{fermion}{o2,v2,i2}
    \fmffreeze
    \fmf{fermion,tension=1.5}{o3,v4,o4}
    \fmf{phantom,tension=1.8}{i4,v4}
    %\fmfv{l.d=20,l.a=180,l={$B_s$\mylbrace{32}{0}}}{B}
    %\fmfv{l.d=20,l.a=0,l={\myrbrace{32}{0}}$X_c$}{X}
    \fmflabel{$q$}{i1}
    \fmflabel{$\bar{b}$}{i2}
    \fmflabel{$\bar{c}(\bar u)$}{o2}
    \fmflabel{$q$}{o1}
    \fmflabel{$\tau^{+}$}{o3}
    \fmflabel{$\nu_\tau$}{o4}
    % boson
    \fmf{boson,label=$W^{+}$,label.side=left}{v4,v2}
    \fmfipair{B,X}
    \fmfiequ{B}{(-.2w,.47h)}
    \fmfiequ{X}{(1.2w,.47h)}
    \fmfiv{l=$B_q\hspace{1pt}\Bigg\{$, l.a=90}{B}
    \fmfiv{l=$\hspace{5pt}\Bigg\}\hspace{1pt}X_q$, l.a=90}{X}
  \end{fmfgraph*}
 \end{fmffile}
 \caption{Leading order Feynman diagram for $B_q \rightarrow X_q\, \tau\nu_\tau$}
 \label{fig:BtoXdiag}
\end{figure}
The invariant transition matrix element for $B_q \to X_q \tau \nu_{\tau}$, where $q$ represent the spectator quarks $q = u,d,s,c$, governing by the quark level transition $b \to c (u) \tau \nu_{\tau}$ as depicted in Fig.~\ref{fig:BtoXdiag}, is expressed as,
%\vspace{-0.2cm}
\begin{equation}
\mathcal{M}(p,k,k_{\tau},k_{\nu_{\tau}})={\frac{\cal G_F}{\sqrt{2}}}V_{c(u)b}\,{\cal H}_\mu(p,k) \,{\cal L}^\mu(k_{\tau},k_{\nu_{\tau}})\,,
\label{eq:TME}
\end{equation}
Here ${\cal G}_F$ is the effective Fermi coupling constant, $V_{c(u)b}$ the relevant CKM matrix element, and ${\cal L}^\mu$ and ${\cal H}_\mu$ denote the leptonic and hadronic currents, respectively. The four-momenta $p, k, k_{\tau}, k_{\nu_{\tau}}$ correspond to the parent meson $B_q = $ ($B$, $B_s$, $B_c$), daughter meson $X = $  \{$D\,(D^*)$, $D_s\,(D_s^*)$, $\eta_c\,(J/\psi)$\}, $\tau$ lepton, and neutrino. In decay processes, mesons are described as momentum-space wave packets representing quark--antiquark momentum and spin distributions. Within the RIQM, the meson bound state $\vert B(\vec{p}, S_B)\rangle$ is expressed as~\cite{Nayak:2021djn},
\begingroup
\setlength{\abovedisplayskip}{6pt}
\setlength{\belowdisplayskip}{6pt}
\setlength{\abovedisplayshortskip}{3pt}
\setlength{\belowdisplayshortskip}{3pt}
\begin{equation}
\begin{aligned}
    \big\vert B (\vec{p},S_{B})\big\rangle 
        &= \hat{\Lambda}(\vec{p},S_{B}) 
        \big\vert (\vec{p_b},\lambda_b);(\vec{p_d},\lambda_d)\big\rangle,\\
    \hat{\Lambda}(\vec{p},S_{B}) 
        &= \frac{\sqrt{3}}{\sqrt{N(\vec{p})}} 
        \sum_{\lambda_b,\lambda_d} \zeta^{B}_{b,d}(\lambda_b,\lambda_d)
        \int d^3p_b\, d^3p_d \; 
        \delta^{(3)}(\vec{p}_b+\vec{p}_d-\vec{p}) \, {\cal G}_{B_s}(\vec{p}_b,\vec{p}_d)
\end{aligned}
\label{eq:MesonWF_aligned}
\end{equation}
 \endgroup
Here, $\sqrt{3}$ is the color factor, and $\zeta^{B}_{b,d}(\lambda_b,\lambda_d)$ the SU(6) spin--flavor coefficient of the $B$ meson. $\vert (\vec{p_b},\lambda_b);(\vec{p_d},\lambda_d)\rangle$ denotes the Fock-space state of a color-singlet quark--antiquark pair with their respective momentum and spin, ${\hat \Lambda}(\vec{p},S_B)$ encoding the meson’s bound-state structure. The same structure can be extended to other meson states: $B_u$, $B_s$, $B_c$, $D$, $D_s$, $\eta_c$, and $J/\psi$. Hadronic matrix elements for transitions between hadrons of definite spin and parity are expressed as tensor structures multiplied by form factors which encode non-perturbative dynamics~\cite{Bernlochner:2021vlv}. 
For $0^- \to 0^-$ and $0^- \to 1^-$ transitions, the processess involve two and four independent form factors, respectively, defined for the vector and axial-vector currents.
%\vspace{-2cm}
\begingroup
\setlength{\abovedisplayskip}{6pt}
\setlength{\belowdisplayskip}{6pt}
\setlength{\abovedisplayshortskip}{3pt}
\setlength{\belowdisplayshortskip}{3pt}
\begin{equation}
\begin{aligned}
    \langle{D\,(D_s)} |\bar c\,\gamma_{\mu}\,b|{B\,(B_s)}\rangle 
        &= (p+k)_{\mu} \, f_{+}(q^2) \,+\, q_{\mu}\,f_{-}(q^2),\\[0.3em]
    \langle{D^*\,(D_s^*)} |\bar c\,\gamma_{\mu}\,b|{B\,(B_s)}\rangle 
        &= \frac{1}{M + m}\, \epsilon^{\sigma^+}\Big \{i\, \epsilon_{\mu\sigma\alpha\beta}\,(p + k)_{\alpha}\,q_{\beta}\,V(q^2)\Big\},\\[0.3em]
    \langle{D^*\,(D_s^*)}| \bar c\,\gamma_{\mu}\,\gamma_{5}\,b|{B\,(B_s)}\rangle 
        &= \frac{1}{M + m}\, \epsilon^{\sigma^+} \Big\{
            g_{\mu \sigma}\, (p+k)_{\mu}\,q_{\mu}\, A_{0}(q^2) \,+\\
        &\quad (p+k)_{\mu}\, (p+k)_{\sigma}\,A_{+}(q^2) \,+\\
        &\quad q_{\mu}\,(p+k)_{\sigma}\,A_{-}(q^2)
        \Big\}.
\end{aligned}
\label{eq:all_ff}
\end{equation}
\endgroup
%\vspace{-3cm}
The hadronic amplitude $\mathcal{H}_{\mu}$ in~\eqref{eq:TME} is derived by evaluating the overlap integral of the meson wavefunctions as in~\eqref{eq:MesonWF_aligned} and in the parent meson rest frame, obtained as,
%\vspace{-3cm}
\begingroup
\setlength{\abovedisplayskip}{6pt}
\setlength{\belowdisplayskip}{6pt}
\setlength{\abovedisplayshortskip}{3pt}
\setlength{\belowdisplayshortskip}{3pt}
\begin{equation}
	{\cal H}_\mu=\sqrt{\frac{ME_k}{N_{B}(0)N_X(\vec{k})}}\int\frac{d^3p_b}{\sqrt{E_{p_b}E_{k+p_b}}}\,{\cal G}_{B}(\vec{p_b},-\vec{p_d})\,{\cal G}_X(\vec{k}+\vec{p_b},-\vec{p_d})\,\langle S_X\vert J_\mu^h(0)\vert S_{B}\rangle. 
    \label{eq:HA}
\end{equation} 	  
\endgroup
%\vspace{-2cm}
Here, $E_{p_b}$ and $E_{k + p_b}$ are the energies of the non-spectator quark in the parent and daughter mesons, respectively, and 
$\langle S_X \vert J_\mu^h(0) \vert S_{B} \rangle$ denotes the spin matrix elements of the hadronic vector–axial vector current. For $0^- \to 0^-$ and $0^- \to 1^-$ transitions, these matrix elements are independently obtained from the RIQM and after comparing with the covariant expansion of form factors Eq.~\eqref{eq:all_ff}, 
yields the model expressions for the form factors $f_{\pm}(q^2)$, $V(q^2)$, $A_0(q^2)$, $A_+(q^2)$, and $A_-(q^2)$. 
\begin{equation}
\begin{aligned}
    f_\pm(q^2) &= \frac{1}{2M} \sqrt{\frac{ME_k}{N_B(0) N_X(\vec{k})}} 
        \int d\vec{p_b}\, {\cal G}_B(\vec{p_b},-\vec{p_d})\, {\cal G}_X(\vec{k}+\vec{p_b},-\vec{p_d})\\
        &\quad \times \frac{(E_{p_b}+m_b)(E_{p_c}+m_c) + |\vec{p_b}|^2 \pm (E_{p_b}+m_b)(M \mp E_k)}
        {E_{p_b} E_{p_c} (E_{p_b}+m_b)(E_{p_c}+m_c)},\\[0.5em]
    V(q^2) &= \frac{M+m}{2M} \sqrt{\frac{ME_k}{N_B(0) N_X(\vec{k})}}
        \int d\vec{p_b}\, {\cal G}_B(\vec{p_b},-\vec{p_d})\, {\cal G}_X(\vec{k}+\vec{p_b},-\vec{p_d})\\
        &\quad \times \sqrt{\frac{E_{p_b}+m_b}{E_{p_b} E_{p_c} (E_{p_c}+m_c)}},\\[0.5em]
    A_0(q^2) &= \frac{1}{M-m} \sqrt{\frac{Mm}{N_B(0) N_X(\vec{k})}}
        \int d\vec{p_b}\, {\cal G}_B(\vec{p_b},-\vec{p_d})\, {\cal G}_X(\vec{k}+\vec{p_b},-\vec{p_d})\\
        &\quad \times \frac{(E_{p_b}+m_b)(E^0_{p_c}+m_c) - \frac{|\vec{p_b}|^2}{3}}
        {\sqrt{E_{p_b} E_{p_c} (E_{p_b}+m_b)(E_{p_c}+m_c)}},\\[0.5em]
    A_\pm(q^2) &= \frac{-E_k (M+m)}{2M(M+2E_k)} 
        \Big[ T \mp \frac{3(M \mp E_k)}{E_k^2 - m^2} \{ I - A_0 (M-m) \} \Big],\\
    &\quad T = J - \frac{M-m}{E_k} A_0,\\[0.3em]
    J &= \sqrt{\frac{ME_k}{N_B(0) N_X(\vec{k})}}
        \int d\vec{p_b}\, {\cal G}_B(\vec{p_b},-\vec{p_d})\, {\cal G}_X(\vec{k}+\vec{p_b},-\vec{p_d})
        \sqrt{\frac{E_{p_b}+m_b}{E_{p_b} E_{p_c} (E_{p_c}+m_c)}},\\
    I &= \sqrt{\frac{ME_k}{N_B(0) N_X(\vec{k})}}
        \int d\vec{p_b}\, {\cal G}_B(\vec{p_b},-\vec{p_d})\, {\cal G}_X(\vec{k}+\vec{p_b},-\vec{p_d})\\
        &\quad \times \frac{(E_{p_b}+m_b)(E^0_{p_c}+m_c)-\frac{|\vec{p_b}|^2}{3}}
        {\sqrt{E_{p_b} E_{p_c} (E_{p_b}+m_b)(E^0_{p_c}+m_c)}},\\
    &\text{where } E^0_{p_c} = \sqrt{|\vec{p_c}|^2 + m_c^2}.
\end{aligned}
\label{eq:formfactors_aligned}
\end{equation}
With the relevant form factors thus obtained in terms of model quantities, the helicity amplitudes and hence the semitauonic decay rates for $B \to D^{(*)}$, $B_s \to D_s^{(*)}$, $B_c \to \eta_c (J/\psi)$ and $B_c \to D^{(*)}$ are evaluated in the following section and our predictions are listed in section~\ref{sec:num_res}.
\subsection{Helicity amplitudes \& decay distribution}
\label{sub:helicity}
Using the weak form factors~\eqref{eq:formfactors_aligned}, the angular decay distribution in $q^2$ is expressed as,
\begin{equation}
\frac{d\Gamma}{dq^2 d\cos\theta} = \frac{{\cal G}_F^2}{(2\pi)^3} |V_{c(u)b}|^2 \frac{(q^2-m_\tau^2)^2}{8M^2 q^2} |\vec{k}| \, {\cal L}^{\mu\sigma} {\cal H}_{\mu\sigma}, 
\quad 0 \le q^2 \le (M-m)^2,
\label{eq:10_compact}
\end{equation}
where \( q = p - k = k_\tau + k_{\nu_\tau} \). \( M \), \(m\) and \( m_\tau \) denote the parent \& daughter meson and \(\tau\)-lepton masses, respectively. For convenience the lepton–hadron structure tensors are written in helicity space after utilizing the completeness property, 
\begin{align}
{\cal L}^{\mu\sigma} {\cal H}_{\mu\sigma} &= {\cal L}_{\mu'\sigma'} g^{\mu'\mu} g^{\sigma'\sigma} {\cal H}_{\mu\sigma} 
= L(m,n)\, g_{mm'} g_{nn'} H(m',n'),\\
L(m,n) &= \epsilon^\mu(m)\, \epsilon^{\sigma\dagger}(n)\, {\cal L}_{\mu\sigma}, \quad
H(m,n) = \epsilon^{\mu\dagger}(m)\, \epsilon^\sigma(n)\, {\cal H}_{\mu\sigma}.
\end{align}
The helicity form factors are obtained by projecting the Lorentz invariant form factors onto the helicity states of the final particles. The Lorentz index contraction in Eq.~\eqref{eq:10_compact} is then performed using these helicity amplitudes linking the covariant and helicity formulations~\cite{Nayak:2021djn}. The azimuthal angle \(\chi\) of the lepton pair is omitted and integrating over \(\chi\) simplifies the decay distribution to depend only on \( q^2 \) and the lepton polar angle \(\theta\). The differential partial helicity rates \( d\Gamma_i / dq^2 \) thus reduce to the form:
\begin{equation}
\frac{d\Gamma_i}{dq^2} = \frac{{\cal G}_F^2 |V_{c(u)b}|^2}{(2\pi)^3} \frac{(q^2-m_\tau^2)^2}{12 M^2 q^2} |\vec{k}| H_i.
\label{eq:dw_compact}
\end{equation}
$H_i$ denotes the conventional set of helicity structure functions each defined as a specific linear combination of the helicity components of hadronic tensor involving the respective form factors.
\begin{align}
H_U = H_+^2 + H_-^2,\,\,\,  H_L = H_0^2,\,\,\, H_P = H_+^2 - H_-^2, \,\,\, H_S = 3 H_t^2,\,\,\, H_{SL} = H_t H_0,
\end{align}
With this definition of the helicity components, required relations between the helicity form factors and the Lorentz invariant form factors can be expressed as follows,
\begin{eqnarray}
	{H_t}=&&{\epsilon^{\mu\dagger}}(t){\epsilon^{\alpha\dagger}_2}(0){\cal H}_{\mu \alpha}\nonumber\\
	=&&\frac{1}{(M+m)}\frac{M{\vert \vec{k}\vert}}{m\sqrt{q^2}}\{(p+k) {.} q\ ({-A_0}+{A_+})+{{q^2}{A_-}}\}\nonumber\\
	{H_\pm}=&&{\epsilon^{\mu\dagger}}(\pm){\epsilon^{\alpha\dagger}_2}(\pm){\cal H}_{\mu \alpha}\\
	=&&\frac{1}{(M+m)}\big\{-(p+k) {.}q\ {A_0}\mp 2M\vert\vec{k}\vert V\big\}\nonumber\\
	{H_0}=&&{\epsilon^{\mu\dagger}}(0){\epsilon^{\alpha\dagger}_2}(0){\cal H}_{\mu \alpha}\nonumber\\
	=&&\frac{1}{(M+m)}\frac{1}{2m\sqrt{q^2}}\Big\{-(p+k){.} q ({M^2}-{m^2}-{q^2}){A_0}+4{M^2}\vert\vec{k}\vert^2 {A_+}\Big\} \nonumber
\end{eqnarray}
\section{Results \& Analysis}
\label{sec:num_res}
To accurately describe the present processes under investigation we use the following input parameters as stated in Table~\ref{table:inputs}. 
\begin{table}[hbt!]
\centering
\small
\setlength\tabcolsep{4pt} % reduce column padding
\renewcommand{\arraystretch}{1.5} % slightly tighter row spacing
\begin{tabular}{|c|c|c|c|p{8cm}|}
\hline
\hline
\textbf{Quarks} & \textbf{Mass (GeV)} & \textbf{Binding Energy (GeV)} & & \textbf{Mass: Mesons \& Leptons (MeV/GeV)} \\
\hline
$b$ & 4.77659 & 4.76633 & & $B^0$ (5279.63), $B^+$ (5279.34), $B_s$ (5366.9), $B_c$ (6274.47), $\tau$ (1776.86) \\
$c$ & 1.49276 & 1.57951 & & $D^0$ (1864.84), $D^0*$ (2006.85), $D^-$ (1869.66), $D^-*$ (2010.26), $D_s$ (1968.34), $D_s^*$ (2112.2) \\
$u$ & 0.07875 & 0.47125 & & $\eta_c$ (2983.9), $J/\psi$ (3096.9) \\
$s$ & 0.31575 & 0.59100 & &  \\
\hline
\multicolumn{5}{|p{\textwidth}|}{\textbf{Other inputs}: $(a,V_0) = (0.017166\ {\rm GeV}^3,-0.1375\ {\rm GeV})$, $V_{\rm cb} = 0.0408\pm0.0014$, $V_{\rm ub} = 0.00382\pm0.00020$, $\tau_{B_c} = 0.51\pm0.009$ ps, $\tau_{B_s} = 1.516\pm0.009$ ps} \\
\hline
\hline
\end{tabular}
\caption{Input parameters used in the analysis: quark masses, masses of mesons and $\tau$ lepton, CKM elements, and lifetimes.}
\label{table:inputs}
\end{table}
\begin{figure}[hbt!]
\captionsetup{width=\textwidth}
\center
\begin{subfigure}{.5\textwidth}
  \center
\includegraphics[width=.8\linewidth]{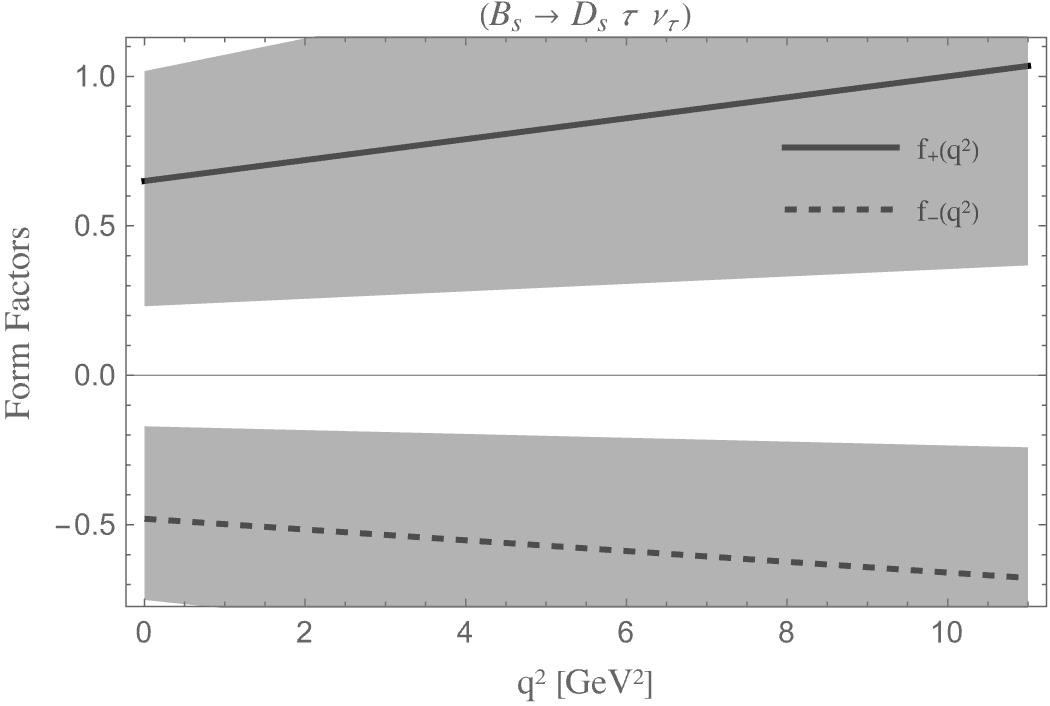}
\end{subfigure}%
\begin{subfigure}{.5\textwidth}
  \center
\includegraphics[width=.8\linewidth]{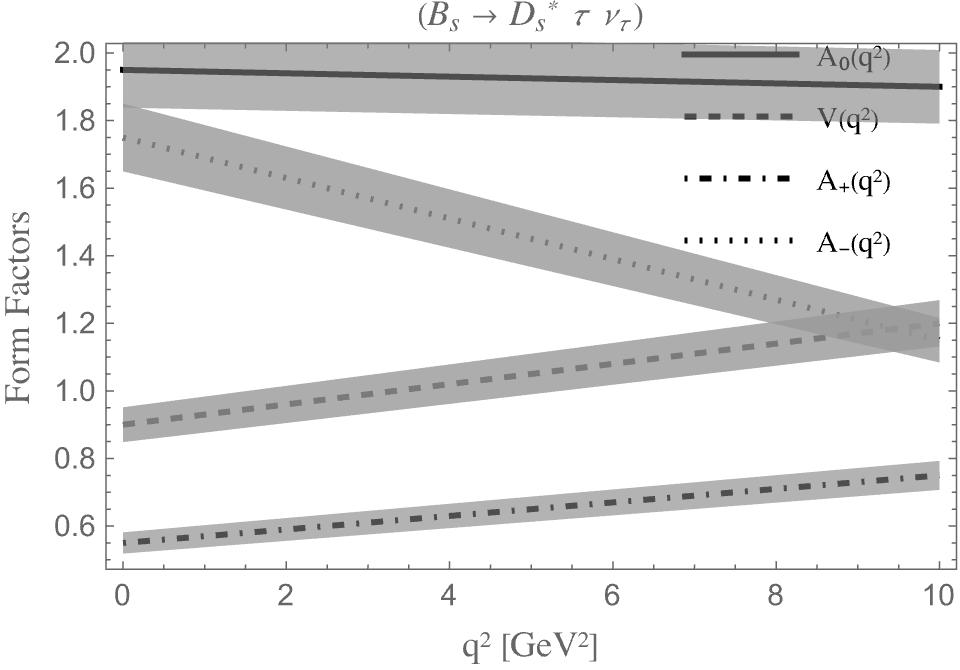}
\end{subfigure}
\begin{subfigure}{.5\textwidth}
  \center
\includegraphics[width=.8\linewidth]{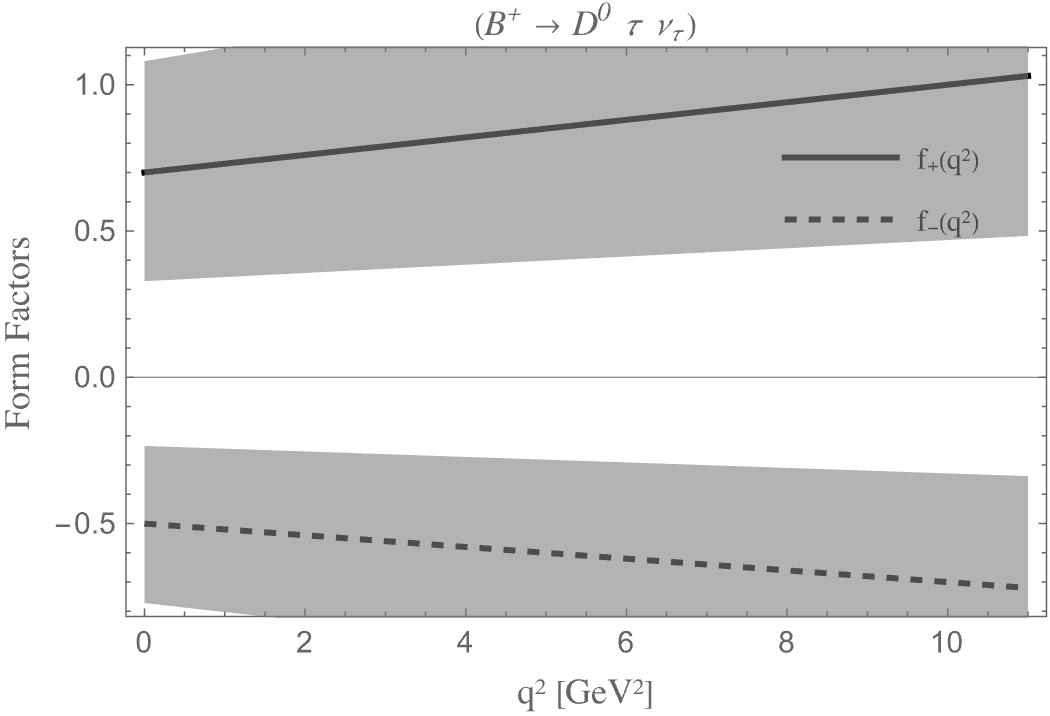}
\end{subfigure}%
\begin{subfigure}{.5\textwidth}
  \center
\includegraphics[width=.8\linewidth]{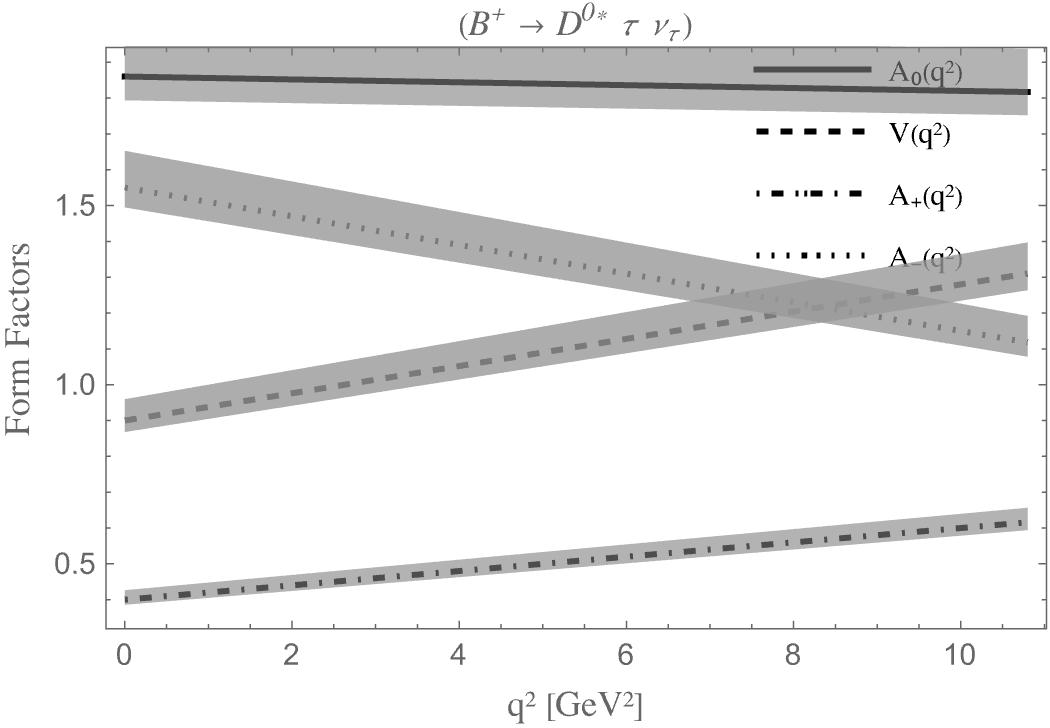}
\end{subfigure}
\caption{$q^2$ distribution spectra of the form factors for $B_s \to D_{s}^{(*)} \tau \nu_{\tau}$ and $B \to D^{(*)} \tau \nu_{\tau}$ transition}
\label{fig:ffgraph}
\end{figure}
With this set of input parameters the Lorentz invariant form factors in~\eqref{eq:formfactors_aligned} are obtained from the overlap integrals of meson wave functions. These form factors depend on the squared momentum transfer 
$q^2 = (p_B - p_X)^2$, varying within $q^2_{\min} = 0$ to $q^2_{\max} = (M_B - m_X)^2$. For $0^- \!\to\! 0^-$ transitions, the dominant contributions arise from $f_+(q^2)$ and $f_{-}(q^2)$ describing vector currents. For $0^- \!\to\! 1^-$ form factors $V(q^2)$, $A_0(q^2)$, $A_+(q^2)$, and $A_-(q^2)$ account for vector and axial-vector components due to spin-1 nature of the final meson. Fig.~\ref{fig:ffgraph} shows the $q^2$-dependence of the form factors for 
$B \!\to\! D^{(*)} \tau \nu_\tau$ and $B_s \!\to\! D_s^{(*)}\tau \nu_\tau$. The form factor $f_+(q^2)$ increases monotonically up to $\sim$11~GeV$^2$ while $f_{-}(q^2)$ decreases with $q^2$ consistent with HQET and pole-dominance expectations~\cite{Wang:2008xt,Fan:2013kqa}. 
The dominance of $f_+(q^2)$ is enhanced for $\tau$ modes due to reduced phase space at large $q^2$. A similar pattern holds for $B_s \!\to\! D_s$, though with slightly larger $f_+(q^2)$ values reflecting internal quark dynamics. For $B \!\to\! D^*$ and $B_s \!\to\! D_s^*$ transitions, $A_0(q^2)$ remains nearly constant, indicating weak $q^2$ dependence of the scalar axial current contribution which governs longitudinal polarization. The $V(q^2)$ and $A_+(q^2)$ form factors rise with recoil while $A_-(q^2)$ decreases due to kinematic suppression. Overall trends align with the covariant light-front quark model~\cite{Verma:2011yw} and are consistent with PQCD~\cite{Hu:2019bdf} and LQCD predictions~\cite{Harrison:2017fmw,Blossier:2021azl,Harrison:2021tol,Aoki:2023qpa} with deviations within model uncertainties. 
Symmetry-breaking effects between $B \!\to\! D^{(*)}$ and $B_s \!\to\! D_s^{(*)}$ remain below 10\%~\cite{Hu:2019bdf}. Compared with BCL~\cite{Fan:2013qz,Boyd:1995sq} or $z$-expansion parameterizations, our RIQ model agrees within 5--10\% at low to moderate $q^2$ with larger deviations near $q^2_{\max}$ due to parametrization sensitivity. A key advantage of our model is its ability to compute form factors over the full $q^2$ range without extrapolation serving inputs to helicity amplitudes for evaluating branching ratios and angular asymmetries outlined below in Table~\ref{Tab:BR},~\ref{Tab:PA}\&~\ref{Tab:PA2}.
\begin{table*}[!htb]
\centering
\resizebox{\textwidth}{!}{%
\begin{tabular}{l|c|c c c|c c c c|c}
\hline \hline
Channels & This work & PQCD~\cite{Hu:2019bdf,Wang:2014yia} & PQCD+Lattice~\cite{Hu:2019bdf} & LQCD~\cite{Harrison:2023dzh,MILC:2015uhg,McLean:2019qcx,Cooper:2021bkt,Harrison:2020gvo} & RQM~\cite{Faustov:2012mt} & RCQM~\cite{Ivanov:2006ni} &LCSR~\cite{Li:2009wq} & HQET~\cite{Fajfer:2012vx} & PDG~\cite{ParticleDataGroup:2024cfk}  \rule[-1ex] {0pt}{1ex} \\
\hline
$B^+ \to D^0 \tau^+ \nu_\tau$ & $0.81_{-0.43}^{+0.44}$ & $0.86^{+0.34}_{-0.25}$ & $0.69^{+0.21}_{-0.17}$ & $0.65 \pm 0.04$ & $-$ & $-$ & $-$ & $0.66\pm0.05$ & $0.77\pm0.25$ \rule[-2ex]{0pt}{1ex}\\
\hline
$B^+ \to D^{0*} \tau^+ \nu_\tau$ & $1.96_{-0.07}^{+0.13}$ & $1.60^{+0.39}_{-0.37}$ & $1.34^{+0.26}_{-0.23}$ & $1.22 \pm 0.07$ & $-$ & $-$ & $-$ & $1.43\pm0.05$ & $1.88\pm0.20$ \rule[-2ex]{0pt}{1ex}\\
\hline
$B^0 \to D^- \tau^+ \nu_\tau$ & $0.75_{-0.41}^{+0.40}$ & $0.82^{+0.33}_{-0.24}$ & $0.62^{+0.19}_{-0.14}$ & $-$ & $-$ & $-$ & $-$ & $0.64\pm0.05$ & $0.99\pm0.21$ \rule[-2ex]{0pt}{1ex}\\
\hline
$B^0 \to D^{*-} \tau^+ \nu_\tau$ & $1.81_{-0.07}^{+0.12}$ & $1.53^{+0.37}_{-0.35}$ & $1.25^{+0.25}_{-0.21}$ & $-$ & $-$ & $-$ & $-$ & $1.29\pm0.06$ & $1.45\pm0.10$ \rule[-2ex]{0pt}{1ex}\\
\hline
$B_s \to D_s \tau \nu_{\tau}$ & $0.76_{-0.49}^{+0.43}$ & $0.72^{+0.32}_{-0.23}$ & $0.63^{+0.17}_{-0.13}$ & $0.74 \pm 0.06$ & $0.62\pm0.05$ & $-$ & $0.33^{+0.14}_{-0.11}$ & $-$ & $-$ \rule[-2ex]{0pt}{1ex}\\
\hline
$B_s \to D_s^* \tau \nu_{\tau}$ & $1.83_{-0.10}^{+0.11}$ & $1.45^{+0.46}_{-0.40}$ & $1.20^{+0.26}_{-0.23}$ & $1.25 \pm 0.05$ & $1.3\pm0.1$ & $-$ & $-$ & $-$ & $-$ \rule[-2ex]{0pt}{1ex}\\
\hline
$B_c \to \eta_c \tau \nu_{\tau}$ & $0.16^{+0.3}_{-0.2}$ & $0.27^{+0.84}_{-0.63}$ & $0.24^{+0.49}_{-0.40}$ & $-$ & $-$ & $0.22$ & $0.26^{+0.6}_{-0.5}$ & $-$ & $-$ \rule[-2ex]{0pt}{1ex}\\
\hline
$B_c \to J/\psi \tau \nu_{\tau}$ & $0.56^{+0.5}_{-0.41}$ & $0.45^{+1.29}_{-1.01}$ & $0.38^{+0.63}_{-0.58}$ & $0.41 (3)$ & $-$ & $0.49$ & $0.53^{+1.6}_{-1.4}$ & $-$ & $-$ \rule[-2ex]{0pt}{1ex}\\
\hline
$B_c \to D \tau \nu_{\tau}$ & $0.0029^{+0.82}_{-0.69}$ & $0.0022^{+0.72}_{-0.52}$ & $-$ & $0.0023(23)$ & $-$ & $0.0021$ & $-$ & $-$ & $-$ \rule[-2ex]{0pt}{1ex}\\
\hline
$B_c \to D^* \tau \nu_{\tau}$ & $0.023^{+0.13}_{-0.12}$ & $0.0064^{+0.20}_{-0.16}$ & $-$ & $-$ & $-$ & $0.0022$ & $-$ & $-$ & $-$ \rule[-2ex]{0pt}{1ex}\\
\hline \hline
\end{tabular}%
}
\caption{RIQM predictions for the branching fractions (in units of $10^{-2}$).}
\label{Tab:BR}
\end{table*}
The total branching fractions are obtained by integrating the differential decay rate~\eqref{eq:dw_compact}. Table~\ref{Tab:BR} lists our predictions (in units of $10^{-2}$) alongside results from PQCD~\cite{Hu:2019bdf},LQCD~\cite{Harrison:2023dzh,MILC:2015uhg,McLean:2019qcx}, LCSR~\cite{Li:2009wq}, HQET~\cite{Fajfer:2012vx}, and quark models~\cite{Faustov:2012mt,Ivanov:2006ni}. Uncertainties include $\pm5\%$ variation from model parameters ($a$, $V_0$) and PDG input errors. Our RIQ model predictions agree well with the available PDG data~\cite{ParticleDataGroup:2024cfk} and remain within theoretical and experimental bounds. For $B^+ \to D^0$ and $B_s \to D_s$ decays, no direct LQCD results exist for absolute branching fractions therefore we estimate them using lattice-predicted ratios $\mathcal{R}_D$ and $\mathcal{R}_{D_s}$~\cite{MILC:2015uhg,McLean:2019qcx}$\,$and experimental $\mathcal{B}(B \to D\,(D_s)\,\ell\nu_\ell)$, finding good consistency, especially for $B_s \to D_s$. Channels with $J^P=1^-$ final states show slightly enhanced branching fractions compared to $J^P=0^+$, reflecting relativistic and spin-structure effects. The close agreement across $B$ and $B_s$ decays implies mild SU(3) breaking ($<10\%$)~\cite{Bordone:2019guc}. 

Using the central value of the $B_c$ meson lifetime, we also determine the branching fractions for the semitauonic $B_c$ decays into charmonium and charm meson final states. Overall the values obtained from Relativistic Constitutent Quark Model~\cite{Ivanov:2006ni} lie within the same order of magnitude. Consistent with the expectations from various theoretical approaches, we find $\mathcal{B}(B_c \to J/\psi)$ is in good agreement with the results reported in Refs.~\cite{Harrison:2020gvo,Ivanov:2006ni}. For $B_c \to \eta_c$ channel our prediction is lower by a factor of $\sim 2$ compared with Refs.~\cite{Ivanov:2006ni} but is consistent with the estimates in~\cite{Ivanov:2006ni}. As anticipated, branching fractions for $B_c \to D^{(*)}$ modes arising from the underlying $b \to u$ transition are significantly smaller than those involving in charmonium final states. Our result for $B_c \to D$ decay are consistent with the predictions reported by LQCD~\cite{Cooper:2021bkt}. For $B_c \to D^*$ transition our values are larger in comparison with Refs.~\cite{Ivanov:2006ni,Wang:2014yia} by a factor of $\sim 5$. Predictions for $B_c \to D^{*}$ and $B_c \to \eta_c$ serve as useful benchmarks in the absence of Lattice data supporting the robustness of the RIQM across pseudoscalar and vector transitions. 
\begin{table*}[!htb]
\centering
\label{tab6}
\vspace{0.2cm}
\begin{tabular}{l| l| c c|cc  } \hline \hline
Observable      & Approach   &  $B^0\to D^- \tau^+\nu_\tau$ & $B^0_s\to D^-_s \tau^+ \nu_\tau $  & $B^0\to D^{*- } \tau^+ \nu_\tau $
& $B^0_s\to D_s^{*-} \tau^+ \nu_l $  \\ \hline
&This Work & $0.293$ & $0.291$ & $-0.53$ & $-0.53$ \\
&LQCD\cite{Harrison:2023dzh} & $-$ & $-$ & $-0.54(19)$ & $-0.53(91)$\\
$P_\tau(D_{(s)}^{(*)})$&PQCD \cite{Hu:2019bdf}&  $0.32(1)$    &$0.31(1)$& $-0.54(1)$& $-0.54(1)$\\
&Belle\cite{Belle:2016dyj}&$-$&$-0.38\pm 0.51^{+0.21}_{-0.16}$&$-$\\   
\hline
&This Work& $-$ & $-$ & $0.48$ & $0.48$\\
&LQCD\cite{Harrison:2023dzh} &$-$ & $-$ & $0.39(24)$ & $0.42(12)$\\
$F_{\rm L}(D_{(s)}^*)$&PQCD \cite{Hu:2019bdf}& $-$&$-$&$0.42(1)$& $0.42(1)$\\
&Belle\cite{Belle:2019ewo}&$-$&$-$&$0.60\pm 0.08\pm 0.04$&$-$\\
&LHCb\cite{LHCb:2023ssl}& $-$ & $-$ & $0.41 \pm 0.06 \pm 0.03$  & $-$\\
\hline\hline
\end{tabular}
\caption{RIQM predictions vs SM and data for $P_\tau(D_{(s)}^{(*)})$ and $F_L(D_{(s)}^*)$}
\label{Tab:PA}
\end{table*}
Using the computed form factors the angular observables: Forward-Backward asymmetry ($\mathcal{A}_{FB}$), asymmetry parameter ($F_L$) which determines the transverse and longitudinal composition of the vector meson final state and longitudinal $\tau$ polarization ($\langle P^{\tau}_{L}\rangle$) for the semitauonic $B$, $B_s$ and $B_c$ decays modes are also evaluated. Table~\ref{Tab:PA} summarizes our results consistent with SM-based predictions and Belle data~\cite{Belle:2016dyj} despite large experimental uncertainties. These observables are far more sensitive to the underlying theoretical framework than the total decay rate. $F_L(D^*) = 0.48$ lies within $2\sigma$ of the Belle and LHCb averages ~\cite{Belle:2019ewo,LHCb:2023ssl} indicating statistical agreement. Their precise measurement can therefore help distinguish between different theoretical models and provide valuable insight into the internal structure and nature of the $B$ meson. These results further validate the reliability of our model in describing angular parameters.  

\begin{figure}[hbt!]
\captionsetup{width=\textwidth}
\center
\begin{subfigure}{.5\textwidth}
  \center
\includegraphics[width=.8\linewidth]{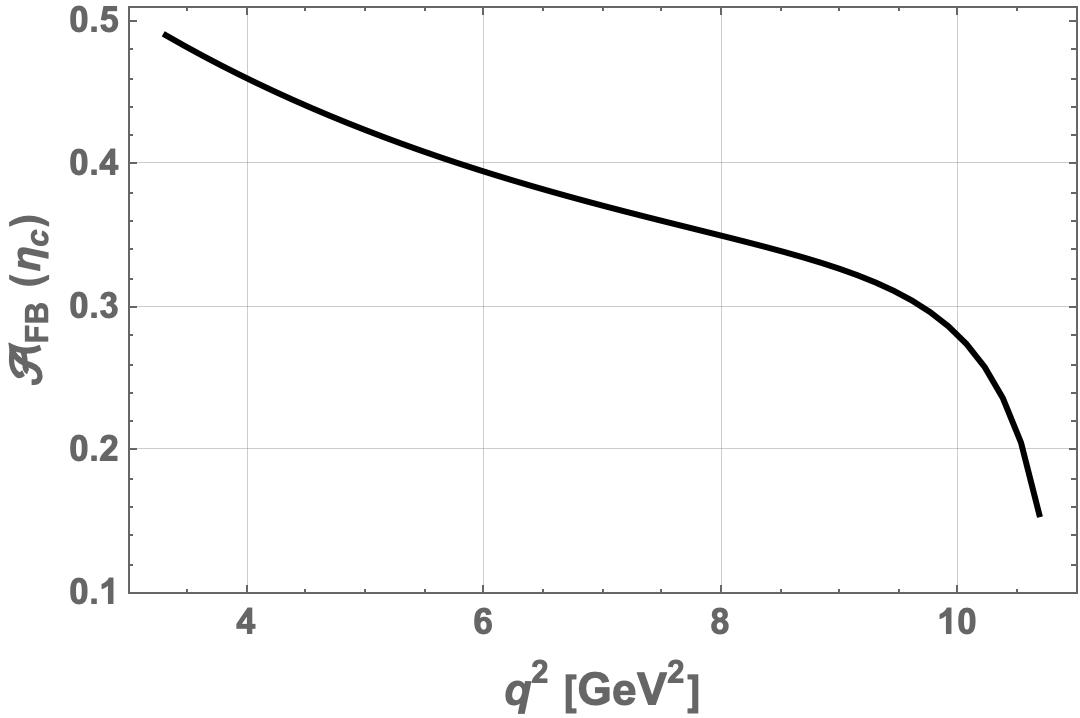}
\end{subfigure}%
\begin{subfigure}{.5\textwidth}
  \center
\includegraphics[width=.8\linewidth]{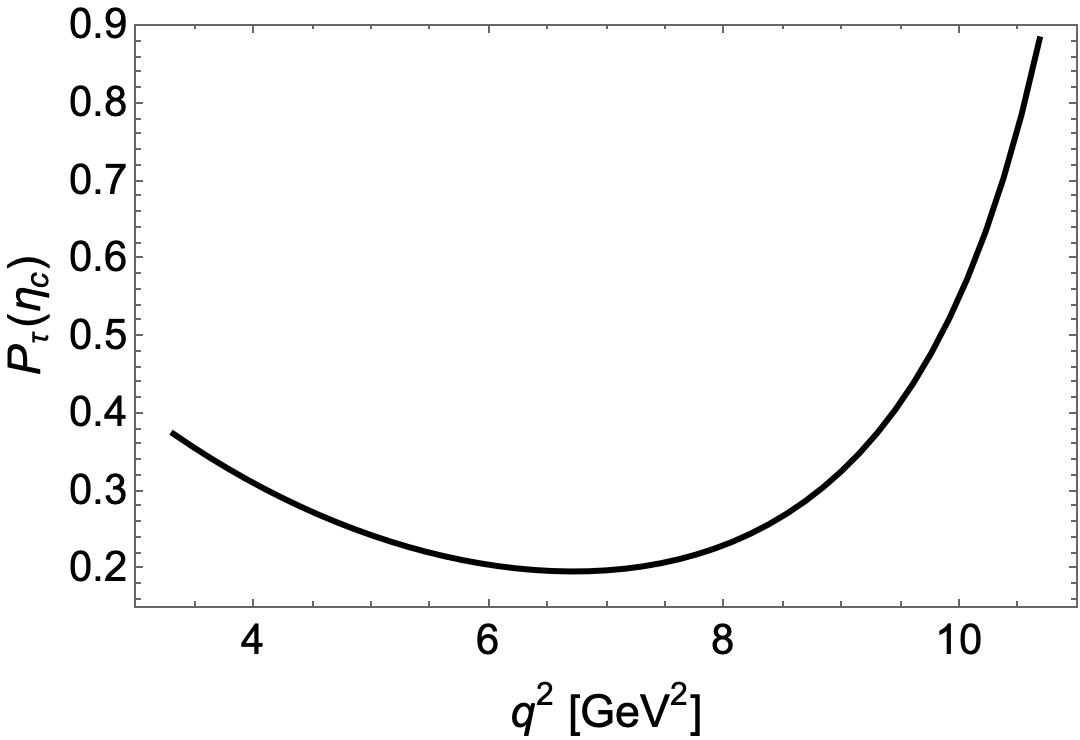}
\end{subfigure}
\begin{subfigure}{.5\textwidth}
  \center
\includegraphics[width=.8\linewidth]{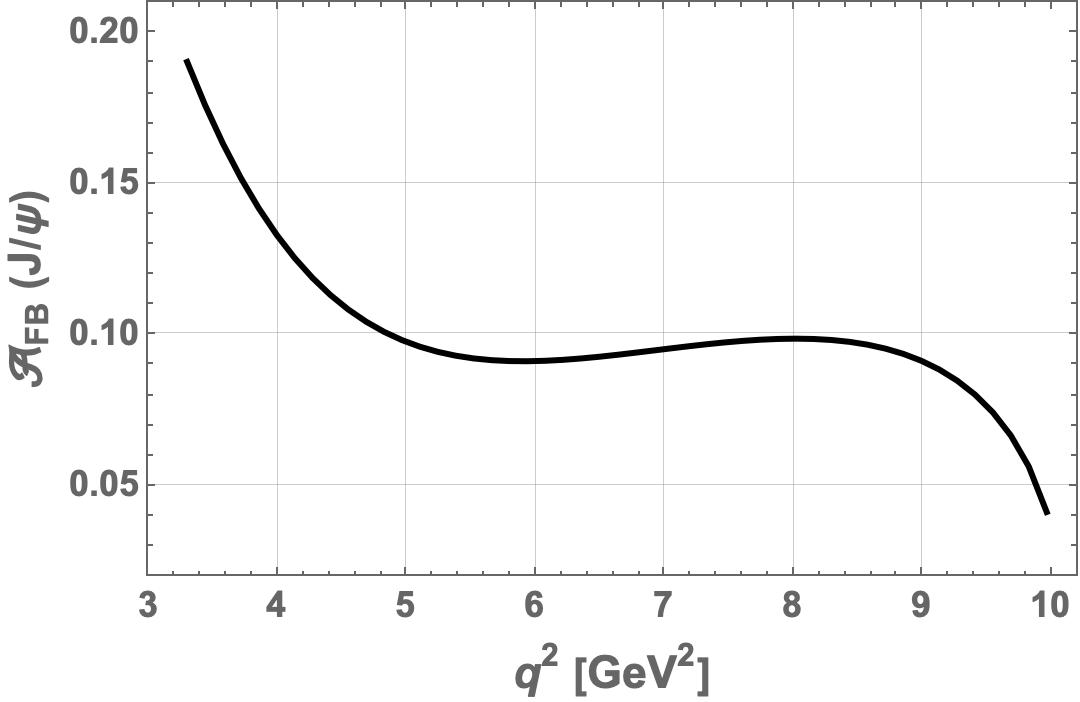}
\end{subfigure}%
\begin{subfigure}{.5\textwidth}
  \center
\includegraphics[width=.8\linewidth]{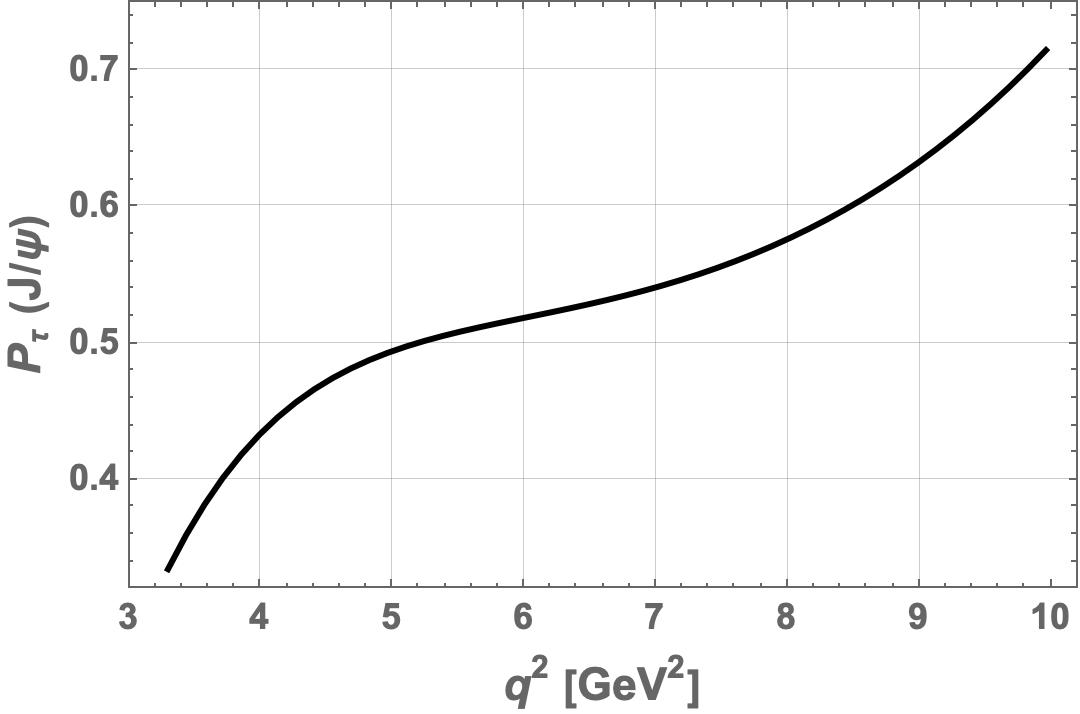}
\end{subfigure}
\begin{subfigure}{.5\textwidth}
  \center
\includegraphics[width=.8\linewidth]{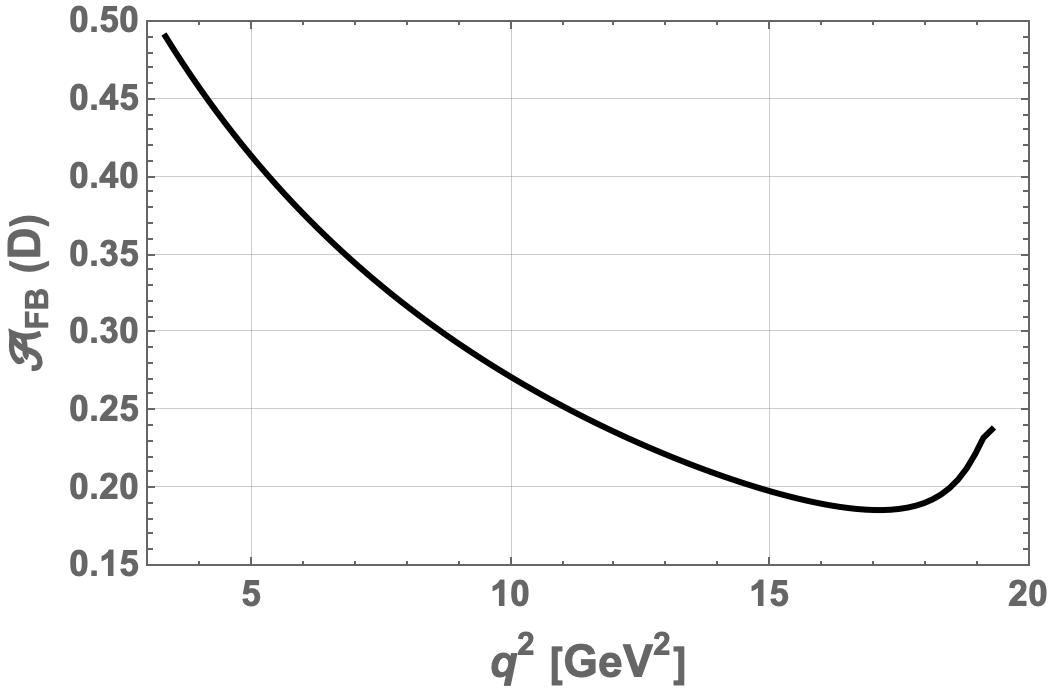}
\end{subfigure}%
\begin{subfigure}{.5\textwidth}
  \center
\includegraphics[width=.8\linewidth]{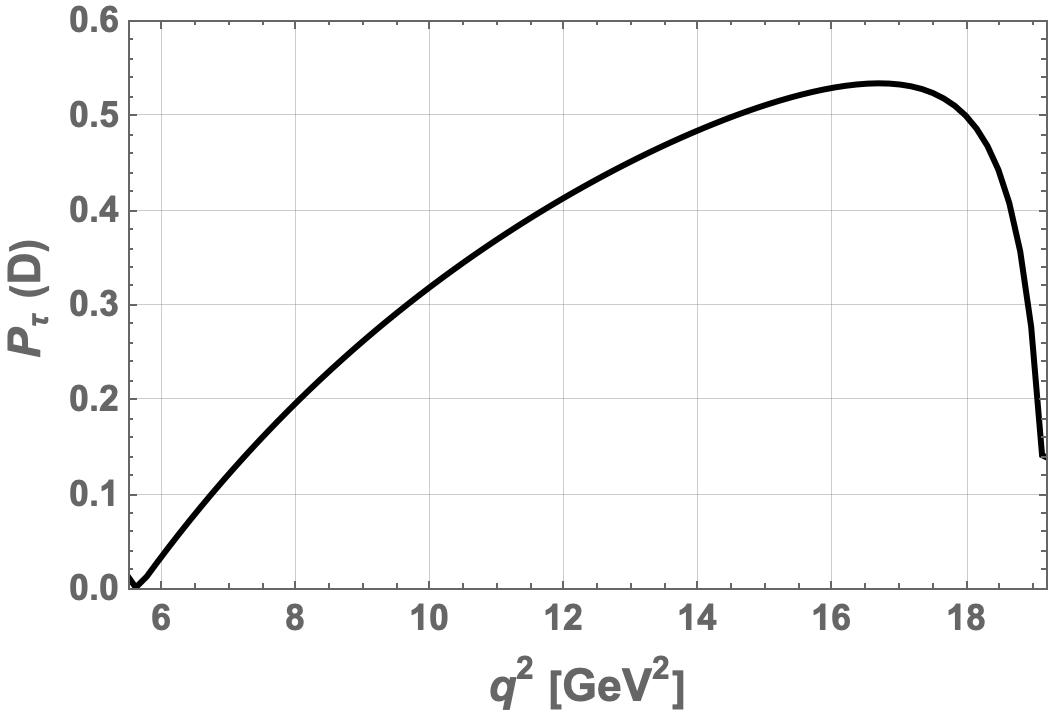}
\end{subfigure}
\begin{subfigure}{.5\textwidth}
  \center
\includegraphics[width=.8\linewidth]{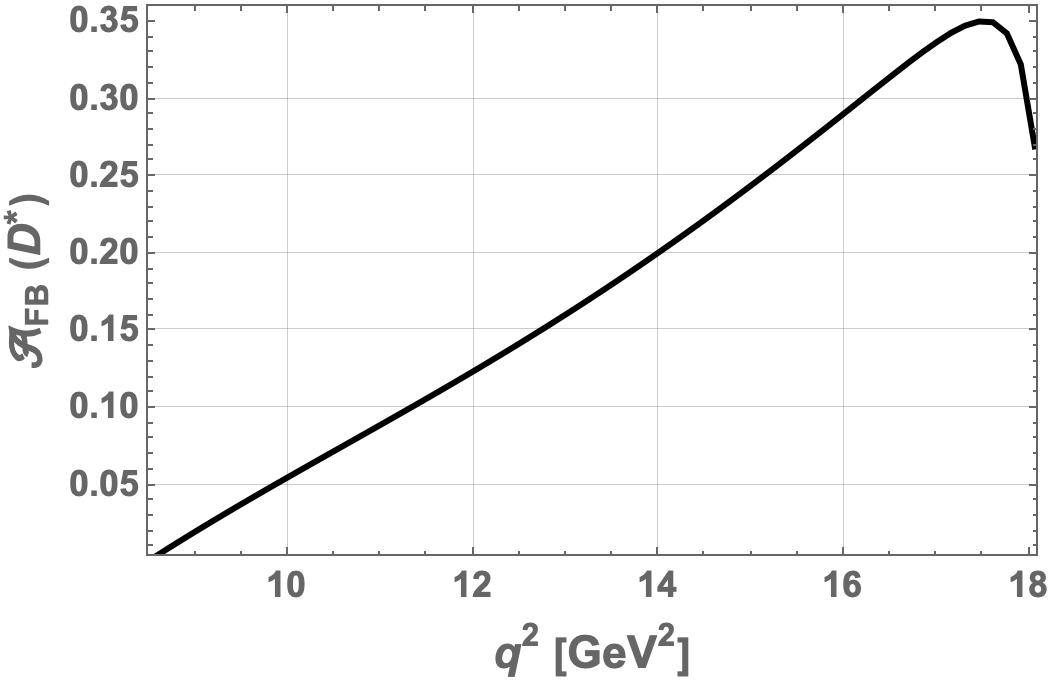}
\end{subfigure}%
\begin{subfigure}{.5\textwidth}
  \center
\includegraphics[width=.8\linewidth]{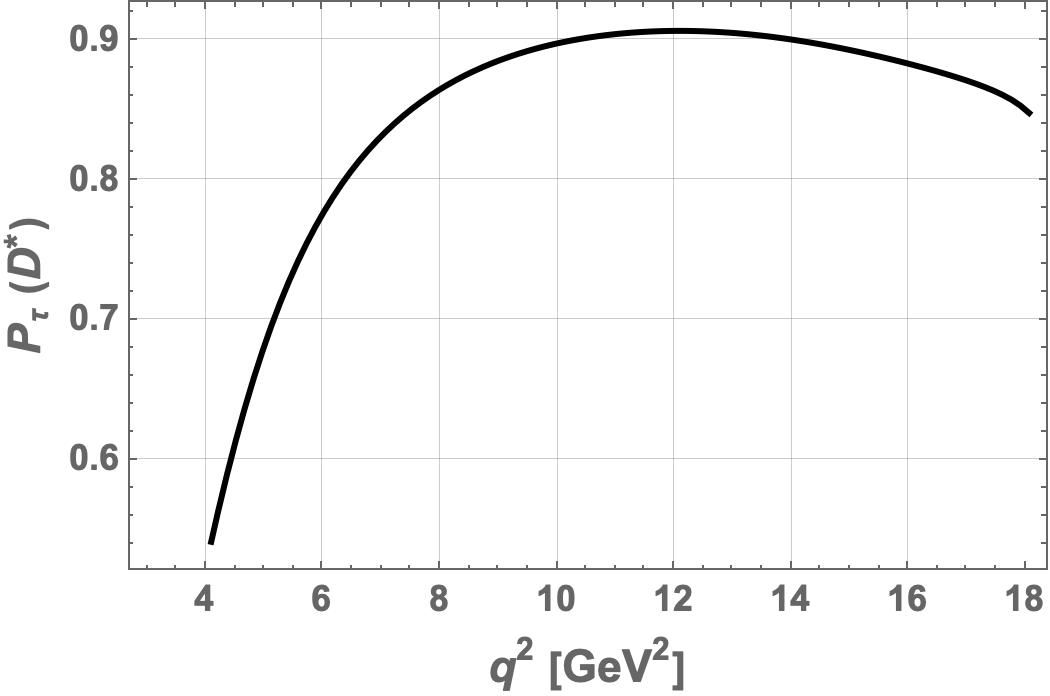}
\end{subfigure}
\caption{$q^2$ distribution spectra of the angular observables in $B_c$ semitauonic decays}
\label{fig:angobs}
\end{figure}
\begin{table}[hbt!]
\centering
\begin{tabular}{|c|c|c|c|c|c|c|}
\hline 
\multirow{2}{*}{Decay process} & 
\multicolumn{3}{c|}{RIQ Model} &
\multicolumn{3}{c|}{Lattice QCD~\cite{Harrison:2020nrv}} \\ \cline{2-7}
& $\mathcal{A}_{FB}$ & $\langle P_{L}^{\tau}\rangle$ & $F_L$ 
& $\mathcal{A}_{FB}$ & $\langle P_{L}^{\tau}\rangle$ & $F_L$ \\ \hline
$B_c\to \eta_c\tau \nu$ & 0.357 & 0.28 & -- & -- & -- & -- \\ \hline
$B_c\to J/\psi \tau\nu$ & 0.093 & 0.56 & -0.066 
& -0.058(12) & 0.5185(75) & 0.4416(92) \\ \hline
$B_c\to D\tau \nu$ & 0.210 & 0.47 & -- & -- & -- & -- \\ \hline
$B_c\to D^*\tau\nu$ & 0.137 & 0.14 & -0.45 & -- & -- & -- \\ \hline
\end{tabular}
\caption{Results of angular observables for $B_c$ decays predicted by RIQM and available LQCD data.}
\label{Tab:PA2}
\end{table}
Table~\ref{Tab:PA2} summarizes our model predictions of angular observables in the $B_c$ decay channels. For transitions involving spin-1 final states, the transverse helicity contribution is significantly larger than the longitudinal one in the $\tau$ mode exceeding it by approximately a factor of two. In the case of the $B_c \to D^*$ transition the dominance of the transverse component becomes even more pronounced with an enhancement of nearly a factor of 7 in the $\tau^-$ mode. To quantify the relative strengths of the transverse and longitudinal polarizations of final vector mesons in the $B_c \to J/\psi$ and $B_c \to D^*$ decays, we evaluate the asymmetry observable $F_L$. For the $B_c \to J/\psi$ channel the resulting value is approximately $-7\%$. In contrast for $B_c \to D^*$ transition our prediction yields a substantially larger negative value, reaching nearly $-45\%$. This pronounced suppression arises from the fact that, in the latter case the scalar‐flip helicity amplitude $\tilde{S}$ becomes the leading contribution and interferes destructively with the remaining amplitudes thereby driving the asymmetry parameter to such a strongly negative value. Our estimate for the $\tau$-polarization $B_c \to J/\psi$ mode, $P_{\tau}(J/\psi)$, shows good agreement with the corresponding LQCD determinations reported in Refs.~\cite{Harrison:2020nrv,Harrison:2020gvo}. Furthermore, the obtained value of $P_{\tau}(\eta_c) = -0.28$ exhibits qualitative compatibility with the results from the PQCD combined with lattice inputs presented in Ref.~\cite{Hu:2019qcn}. In addition, we present predictions of the observables in $B_c \to D$ and $B_c \to D^*$ for which LQCD results are currently not available. Hence our theoretical estimates may serve as a valuable reference for probing angular observables in these charm final-states of $B_c$ semitauonic decays.

Fig.~\ref{fig:angobs} depicts the $q^2$ distributions of forward-backward asymmetry and longitudinal $\tau$-polarization in $B_c \rightarrow (J/\psi, \eta_c, D, D^*)$ decays. For the pseudoscalar transition $B_c \rightarrow \eta_c$ the $\tau$-polarization spectra start near $q^2 \simeq 2~{\rm GeV}^2$ and decrease over $2 \le q^2 \le 8~{\rm GeV}^2$ due to limited phase space. Unlike $B_c \rightarrow D$ which grows smoothly with $q^2$, the quantity $P_{\tau}(\eta_c)$ rises steadily at higher $q^2$. This trend reflects the absence of the parity-odd helicity structure $H_p$ in pseudoscalar amplitudes, leading to strong $q^2$ sensitivity across the spectrum. For $B_c \rightarrow (J/\psi, D^*)$, $P_{\tau}$ shows a rapid increase in the large $q^2$ region driven by the dominant no-flip part of the transverse helicity amplitude. The $q^2$–dependence of $P_{\tau}(J/\psi)$ agrees qualitatively with Lattice predictions~\cite{Harrison:2020nrv}. The $q^2$ distributions of $\mathcal{A}_{\rm FB}$ for $\eta_c$ and $D$, the spectra falls near zero recoil while for $J/\psi$, it decreases at low $q^2$ upto $\sim 6~{\rm GeV}^2$ and then drops sharply at high $q^2$ due to the threshold factor ${(q^2-m_{\tau}^2)}/{q^2}$ suppressing the longitudinal amplitude $L$. The shape of $\mathcal{A}_{\rm FB}(J/\psi)$ is consistent with Lattice results~\cite{Harrison:2020nrv}. For $B_c \rightarrow D^*$, the spin-flip contribution is small but the scalar flip term $\tilde{S}$ becomes relevant at large $q^2$ shaping the overall behavior of $\mathcal{A}_{\rm FB}(D^*)$.
\section{Summary}
\label{sec:conclusions}
The present investigation provide timely theoretical inputs \& outcomes for ongoing and future experiments, offering insights into SM dynamics with continued synergy between phenomenology, Lattice studies and measurements being crucial for probing semileptonic heavy meson decays involving $\tau$ leptons. Further improvements can be made in this analysis by adopting a \textit{correlated approach} where we will constrain the model parameters and extract the meson wavefunctions using LQCD inputs of $B_c \to J/\psi$ form factors. This Lattice-constrained framework will then be employed to compute the full $q^2$ dependence of form factors for a broader class of decay channels. By combining phenomenological modeling with robust non-perturbative inputs, this approach will significantly improve reliability over purely potential model based predictions.
\bibliography{bibliography}{}

\providecommand{\href}[2]{#2}\begingroup\raggedright\begin{thebibliography}{10}

\bibitem{Belle:2016dyj}
{\bfseries Belle} Collaboration, S.~Hirose {\em et~al.}, ``{Measurement of the $\tau$ lepton polarization and $R(D^*)$ in the decay $\bar{B} \rightarrow D^* \tau^- \bar{\nu}_\tau$},'' \href{http://dx.doi.org/10.1103/PhysRevLett.118.211801}{{\em Phys. Rev. Lett.} {\bfseries 118} no.~21, (2017) 211801}, \href{http://arxiv.org/abs/1612.00529}{{\ttfamily arXiv:1612.00529 [hep-ex]}}.

\bibitem{Belle:2019ewo}
{\bfseries Belle} Collaboration, A.~Abdesselam {\em et~al.}, ``{Measurement of the $D^{\ast-}$ polarization in the decay $B^0 \to D^{\ast -}\tau^+\nu_{\tau}$},'' in {\em {10th International Workshop on the CKM Unitarity Triangle}}.
\newblock 3, 2019.
\newblock \href{http://arxiv.org/abs/1903.03102}{{\ttfamily arXiv:1903.03102 [hep-ex]}}.

\bibitem{Huang:2018nnq}
Z.-R. Huang, Y.~Li, C.-D. Lu, M.~A. Paracha, and C.~Wang, ``{Footprints of New Physics in $b\to c\tau\nu$ Transitions},'' \href{http://dx.doi.org/10.1103/PhysRevD.98.095018}{{\em Phys. Rev. D} {\bfseries 98} no.~9, (2018) 095018}, \href{http://arxiv.org/abs/1808.03565}{{\ttfamily arXiv:1808.03565 [hep-ph]}}.

\bibitem{Bhattacharya:2018kig}
S.~Bhattacharya, S.~Nandi, and S.~Kumar~Patra, ``{$b \rightarrow c \tau \nu _{\tau }$ Decays: a catalogue to compare, constrain, and correlate new physics effects},'' \href{http://dx.doi.org/10.1140/epjc/s10052-019-6767-7}{{\em Eur. Phys. J. C} {\bfseries 79} no.~3, (2019) 268}, \href{http://arxiv.org/abs/1805.08222}{{\ttfamily arXiv:1805.08222 [hep-ph]}}.

\bibitem{LHCb:2023ssl}
{\bfseries LHCb} Collaboration, R.~Aaij {\em et~al.}, ``{Measurement of the $D^{*}$ longitudinal polarization in $B^0\to D^{*-}\tau^{+}\nu_{\tau}$ decays},'' \href{http://arxiv.org/abs/2311.05224}{{\ttfamily arXiv:2311.05224 [hep-ex]}}.

\bibitem{Khlopov:1978id}
M.~Y. Khlopov, ``{Effects of Symmetry Violation in Semileptonic Meson Decays},'' {\em Sov. J. Nucl. Phys.} {\bfseries 28} (1978) 583.

\bibitem{Chang:1992pt}
C.-H. Chang and Y.-Q. Chen, ``{The Decays of B(c) meson},'' \href{http://dx.doi.org/10.1103/PhysRevD.49.3399}{{\em Phys. Rev. D} {\bfseries 49} (1994) 3399--3411}.

\bibitem{Dey:2025xdx}
U.~Dey and S.~Nandi, ``{Correlated study on some B$_{c}${\textrightarrow} P and B$_{c}${\textrightarrow} S wave channels in light of new inputs},'' \href{http://dx.doi.org/10.1007/JHEP07(2025)144}{{\em JHEP} {\bfseries 07} (2025) 144}, \href{http://arxiv.org/abs/2503.01693}{{\ttfamily arXiv:2503.01693 [hep-ph]}}.

\bibitem{Bernlochner:2021vlv}
F.~U. Bernlochner, M.~F. Sevilla, D.~J. Robinson, and G.~Wormser, ``{Semitauonic b-hadron decays: A lepton flavor universality laboratory},'' \href{http://dx.doi.org/10.1103/RevModPhys.94.015003}{{\em Rev. Mod. Phys.} {\bfseries 94} no.~1, (2022) 015003}, \href{http://arxiv.org/abs/2101.08326}{{\ttfamily arXiv:2101.08326 [hep-ex]}}.

\bibitem{Nayak:2021djn}
L.~Nayak, S.~Patnaik, P.~C. Dash, S.~Kar, and N.~Barik, ``{Lepton mass effects in exclusive semileptonic $B_c$-meson decays},'' \href{http://dx.doi.org/10.1103/PhysRevD.104.036012}{{\em Phys. Rev. D} {\bfseries 104} (2021) 036012}, \href{http://arxiv.org/abs/2106.09463}{{\ttfamily arXiv:2106.09463 [hep-ph]}}.

\bibitem{Nayak:2024esq}
L.~Nayak, S.~Patnaik, P.~Sadangi, and S.~K. Swain, ``{Study of exclusive decays of Bs\textrightarrow{}\ensuremath{\psi}(1S,2S)Ks and Bs\textrightarrow{}\ensuremath{\eta}c(1S,2S)Ks},'' \href{http://dx.doi.org/10.1103/PhysRevD.110.113003}{{\em Phys. Rev. D} {\bfseries 110} no.~11, (2024) 113003}, \href{http://arxiv.org/abs/2404.14267}{{\ttfamily arXiv:2404.14267 [hep-ph]}}.

\bibitem{Patnaik:2017cbl}
S.~Patnaik, P.~C. Dash, S.~Kar, S.~Patra, and N.~Barik, ``{Magnetic dipole transitions of $B_c$ and $B_c^*$ mesons in the relativistic independent quark model},'' \href{http://dx.doi.org/10.1103/PhysRevD.96.116010}{{\em Phys. Rev. D} {\bfseries 96} no.~11, (2017) 116010}, \href{http://arxiv.org/abs/1710.08242}{{\ttfamily arXiv:1710.08242 [hep-ph]}}. [Erratum: Phys.Rev.D 99, 019901 (2019)].

\bibitem{Patnaik:2023efe}
S.~Patnaik, L.~Nayak, P.~Sadangi, S.~Swain, and R.~Singh, ``{Study of angular observables in exclusive semileptonic Bc decays},'' \href{http://dx.doi.org/10.1103/PhysRevD.110.055028}{{\em Phys. Rev. D} {\bfseries 110} no.~5, (2024) 055028}, \href{http://arxiv.org/abs/2312.17114}{{\ttfamily arXiv:2312.17114 [hep-ph]}}.

\bibitem{Patnaik:2025fry}
S.~Patnaik, L.~Nayak, and S.~K. Swain, ``{B{\textrightarrow}D(*){\ensuremath{\tau}}{\ensuremath{\nu}}{\ensuremath{\tau}} decay properties with the relativistic independent quark model},'' \href{http://dx.doi.org/10.1103/xtcf-2vw9}{{\em Phys. Rev. D} {\bfseries 112} no.~3, (2025) 033003}.

\bibitem{Barik:1986mq}
N.~Barik and B.~K. Dash, ``{Mass Spectrum of Low Lying Baryons in the Ground State in a Relativistic Potential Model of Independent Quarks With Chiral Symmetry},'' \href{http://dx.doi.org/10.1103/PhysRevD.33.1925}{{\em Phys. Rev. D} {\bfseries 33} (1986) 1925--1933}.

\bibitem{Barik:1987zb}
N.~Barik, B.~K. Dash, and P.~C. Dash, ``{The ($Q \bar{Q}$) Pion and Its Decay Constant in a Chiral Potential Model},'' \href{http://dx.doi.org/10.1007/BF02845835}{{\em Pramana} {\bfseries 29} (1987) 543--557}.

\bibitem{Wong:1980ce}
C.~W. Wong, ``{Center-of-mass Correction in the {MIT} Bag Model. 2.}'' \href{http://dx.doi.org/10.1103/PhysRevD.24.1416}{{\em Phys. Rev. D} {\bfseries 24} (1981) 1416}.

\bibitem{Nayak:2025amk}
L.~Nayak, S.~Patnaik, D.~Pandey, and S.~K. Swain, ``{Mass spectrum of S-wave mesons in the relativistic independent quark model},'' \href{http://arxiv.org/abs/2504.05714}{{\ttfamily arXiv:2504.05714 [hep-ph]}}.

\bibitem{Dash:2023ohf}
K.~Dash, P.~C. Dash, R.~N. Panda, S.~Kar, and N.~Barik, ``{Purely leptonic decays of heavy-flavored charged mesons},'' \href{http://dx.doi.org/10.1103/PhysRevD.110.053004}{{\em Phys. Rev. D} {\bfseries 110} no.~5, (2024) 053004}, \href{http://arxiv.org/abs/2312.06130}{{\ttfamily arXiv:2312.06130 [hep-ph]}}.

\bibitem{Wang:2008xt}
W.~Wang, Y.-L. Shen, and C.-D. Lu, ``{Covariant Light-Front Approach for B(c) transition form factors},'' \href{http://dx.doi.org/10.1103/PhysRevD.79.054012}{{\em Phys. Rev. D} {\bfseries 79} (2009) 054012}, \href{http://arxiv.org/abs/0811.3748}{{\ttfamily arXiv:0811.3748 [hep-ph]}}.

\bibitem{Fan:2013kqa}
Y.-Y. Fan, W.-F. Wang, and Z.-J. Xiao, ``{Study of $\bar{B}_s^0 \to (D_s^+,D_s^{*+}) l^-\bar{\nu}_l$ decays in the pQCD factorization approach},'' \href{http://dx.doi.org/10.1103/PhysRevD.89.014030}{{\em Phys. Rev. D} {\bfseries 89} no.~1, (2014) 014030}, \href{http://arxiv.org/abs/1311.4965}{{\ttfamily arXiv:1311.4965 [hep-ph]}}.

\bibitem{Verma:2011yw}
R.~C. Verma, ``{Decay constants and form factors of s-wave and p-wave mesons in the covariant light-front quark model},'' \href{http://dx.doi.org/10.1088/0954-3899/39/2/025005}{{\em J. Phys. G} {\bfseries 39} (2012) 025005}, \href{http://arxiv.org/abs/1103.2973}{{\ttfamily arXiv:1103.2973 [hep-ph]}}.

\bibitem{Hu:2019bdf}
X.-Q. Hu, S.-P. Jin, and Z.-J. Xiao, ``{Semileptonic decays $B/B_s \to (D^{(*)},D_s^{(*)}) l \nu_l$ in the PQCD approach with the lattice QCD input},'' \href{http://dx.doi.org/10.1088/1674-1137/44/5/053102}{{\em Chin. Phys. C} {\bfseries 44} no.~5, (2020) 053102}, \href{http://arxiv.org/abs/1912.03981}{{\ttfamily arXiv:1912.03981 [hep-ph]}}.

\bibitem{Harrison:2017fmw}
{\bfseries HPQCD} Collaboration, J.~Harrison, C.~Davies, and M.~Wingate, ``{Lattice QCD calculation of the ${{B}_{(s)}\to D_{(s)}^{*}\ell{\nu}}$ form factors at zero recoil and implications for ${|V_{cb}|}$},'' \href{http://dx.doi.org/10.1103/PhysRevD.97.054502}{{\em Phys. Rev. D} {\bfseries 97} no.~5, (2018) 054502}, \href{http://arxiv.org/abs/1711.11013}{{\ttfamily arXiv:1711.11013 [hep-lat]}}.

\bibitem{Blossier:2021azl}
B.~Blossier, P.~H. Cahue, J.~Heitger, S.~La~Cesa, J.~Neuendorf, and S.~Zafeiropoulos, ``{$B_s \to D^{(*)}_s$ form factors from lattice QCD with ${\rm N_f}=2$ Wilson-clover quarks},'' \href{http://dx.doi.org/10.22323/1.396.0056}{{\em PoS} {\bfseries LATTICE2021} (2022) 056}, \href{http://arxiv.org/abs/2111.05733}{{\ttfamily arXiv:2111.05733 [hep-lat]}}.

\bibitem{Harrison:2021tol}
{\bfseries HPQCD} Collaboration, J.~Harrison and C.~T.~H. Davies, ``{Bs\textrightarrow{}Ds* form factors for the full q2 range from lattice QCD},'' \href{http://dx.doi.org/10.1103/PhysRevD.105.094506}{{\em Phys. Rev. D} {\bfseries 105} no.~9, (2022) 094506}, \href{http://arxiv.org/abs/2105.11433}{{\ttfamily arXiv:2105.11433 [hep-lat]}}.

\bibitem{Aoki:2023qpa}
{\bfseries JLQCD} Collaboration, Y.~Aoki, B.~Colquhoun, H.~Fukaya, S.~Hashimoto, T.~Kaneko, R.~Kellermann, J.~Koponen, and E.~Kou, ``{B\textrightarrow{}D*\ensuremath{\ell}\ensuremath{\nu}\ensuremath{\ell} semileptonic form factors from lattice QCD with M\"obius domain-wall quarks},'' \href{http://dx.doi.org/10.1103/PhysRevD.109.074503}{{\em Phys. Rev. D} {\bfseries 109} no.~7, (2024) 074503}, \href{http://arxiv.org/abs/2306.05657}{{\ttfamily arXiv:2306.05657 [hep-lat]}}.

\bibitem{Fan:2013qz}
Y.-Y. Fan, W.-F. Wang, S.~Cheng, and Z.-J. Xiao, ``{Semileptonic decays $B \to D^{(*)} l\nu$ in the perturbative QCD factorization approach},'' \href{http://dx.doi.org/10.1007/s11434-013-0049-9}{{\em Chin. Sci. Bull.} {\bfseries 59} (2014) 125--132}, \href{http://arxiv.org/abs/1301.6246}{{\ttfamily arXiv:1301.6246 [hep-ph]}}.

\bibitem{Boyd:1995sq}
C.~G. Boyd, B.~Grinstein, and R.~F. Lebed, ``{Model independent determinations of anti-B ---\ensuremath{>} D (lepton), D* (lepton) anti-neutrino form-factors},'' \href{http://dx.doi.org/10.1016/0550-3213(95)00653-2}{{\em Nucl. Phys. B} {\bfseries 461} (1996) 493--511}, \href{http://arxiv.org/abs/hep-ph/9508211}{{\ttfamily arXiv:hep-ph/9508211}}.

\bibitem{Wang:2014yia}
W.-F. Wang, X.~Yu, C.-D. L{\"u}, and Z.-J. Xiao, ``{Semileptonic decays $B_c^+$ {\textrightarrow} $D_{(s)}^{(*)}(l^+ν_l,l^+l^-,ν\bar{ν}$) in the perturbative QCD approach},'' \href{http://dx.doi.org/10.1103/PhysRevD.90.094018}{{\em Phys. Rev. D} {\bfseries 90} no.~9, (2014) 094018}, \href{http://arxiv.org/abs/1401.0391}{{\ttfamily arXiv:1401.0391 [hep-ph]}}.

\bibitem{Harrison:2023dzh}
{\bfseries HPQCD, (HPQCD Collaboration)\textdaggerdbl{}} Collaboration, J.~Harrison and C.~T.~H. Davies, ``{B\textrightarrow{}D* and Bs\textrightarrow{}Ds* vector, axial-vector and tensor form factors for the full q2 range from lattice QCD},'' \href{http://dx.doi.org/10.1103/PhysRevD.109.094515}{{\em Phys. Rev. D} {\bfseries 109} no.~9, (2024) 094515}, \href{http://arxiv.org/abs/2304.03137}{{\ttfamily arXiv:2304.03137 [hep-lat]}}.

\bibitem{MILC:2015uhg}
{\bfseries MILC} Collaboration, J.~A. Bailey {\em et~al.}, ``{B\textrightarrow{}D\ensuremath{\ell}\ensuremath{\nu} form factors at nonzero recoil and |V$_{cb}$| from 2+1-flavor lattice QCD},'' \href{http://dx.doi.org/10.1103/PhysRevD.92.034506}{{\em Phys. Rev. D} {\bfseries 92} no.~3, (2015) 034506}, \href{http://arxiv.org/abs/1503.07237}{{\ttfamily arXiv:1503.07237 [hep-lat]}}.

\bibitem{McLean:2019qcx}
E.~McLean, C.~T.~H. Davies, J.~Koponen, and A.~T. Lytle, ``{$B_s\to D_s \ell\nu$ Form Factors for the full $q^2$ range from Lattice QCD with non-perturbatively normalized currents},'' \href{http://dx.doi.org/10.1103/PhysRevD.101.074513}{{\em Phys. Rev. D} {\bfseries 101} no.~7, (2020) 074513}, \href{http://arxiv.org/abs/1906.00701}{{\ttfamily arXiv:1906.00701 [hep-lat]}}.

\bibitem{Cooper:2021bkt}
{\bfseries HPQCD} Collaboration, L.~J. Cooper, C.~T.~H. Davies, and M.~Wingate, ``{Form factors for the processes $B^+_c \to D^0 \ell^+\nu_\ell$ and $B^+_c \to D^+_s \ell^+ \ell^+ (\nu \bar \nu)$ from lattice QCD},'' \href{http://dx.doi.org/10.1103/PhysRevD.105.014503}{{\em Phys. Rev. D} {\bfseries 105} no.~1, (2022) 014503}, \href{http://arxiv.org/abs/2108.11242}{{\ttfamily arXiv:2108.11242 [hep-lat]}}.

\bibitem{Harrison:2020gvo}
{\bfseries HPQCD} Collaboration, J.~Harrison, C.~T.~H. Davies, and A.~Lytle, ``{$B_c \rightarrow J/\psi$ form factors for the full $q^2$ range from lattice QCD},'' \href{http://dx.doi.org/10.1103/PhysRevD.102.094518}{{\em Phys. Rev. D} {\bfseries 102} no.~9, (2020) 094518}, \href{http://arxiv.org/abs/2007.06957}{{\ttfamily arXiv:2007.06957 [hep-lat]}}.

\bibitem{Faustov:2012mt}
R.~N. Faustov and V.~O. Galkin, ``{Weak decays of $B_s$ mesons to $D_s$ mesons in the relativistic quark model},'' \href{http://dx.doi.org/10.1103/PhysRevD.87.034033}{{\em Phys. Rev. D} {\bfseries 87} no.~3, (2013) 034033}, \href{http://arxiv.org/abs/1212.3167}{{\ttfamily arXiv:1212.3167 [hep-ph]}}.

\bibitem{Ivanov:2006ni}
M.~A. Ivanov, J.~G. Korner, and P.~Santorelli, ``{Exclusive semileptonic and nonleptonic decays of the $B_c$ meson},'' \href{http://dx.doi.org/10.1103/PhysRevD.73.054024}{{\em Phys. Rev. D} {\bfseries 73} (2006) 054024}, \href{http://arxiv.org/abs/hep-ph/0602050}{{\ttfamily arXiv:hep-ph/0602050}}.

\bibitem{Li:2009wq}
R.-H. Li, C.-D. Lu, and Y.-M. Wang, ``{Exclusive $B_s$ decays to the charmed mesons $D^+_{s}$(1968,2317) in the standard model},'' \href{http://dx.doi.org/10.1103/PhysRevD.80.014005}{{\em Phys. Rev. D} {\bfseries 80} (2009) 014005}, \href{http://arxiv.org/abs/0905.3259}{{\ttfamily arXiv:0905.3259 [hep-ph]}}.

\bibitem{Fajfer:2012vx}
S.~Fajfer, J.~F. Kamenik, and I.~Nisandzic, ``{On the $B \to D^* \tau \bar \nu_{\tau}$ Sensitivity to New Physics},'' \href{http://dx.doi.org/10.1103/PhysRevD.85.094025}{{\em Phys. Rev. D} {\bfseries 85} (2012) 094025}, \href{http://arxiv.org/abs/1203.2654}{{\ttfamily arXiv:1203.2654 [hep-ph]}}.

\bibitem{ParticleDataGroup:2024cfk}
{\bfseries Particle Data Group} Collaboration, S.~Navas {\em et~al.}, ``{Review of particle physics},'' \href{http://dx.doi.org/10.1103/PhysRevD.110.030001}{{\em Phys. Rev. D} {\bfseries 110} no.~3, (2024) 030001}.

\bibitem{Bordone:2019guc}
M.~Bordone, N.~Gubernari, D.~van Dyk, and M.~Jung, ``{Heavy-Quark expansion for ${{\bar{B}}_s\rightarrow D^{(*)}_s}$ form factors and unitarity bounds beyond the ${SU(3)_F}$ limit},'' \href{http://dx.doi.org/10.1140/epjc/s10052-020-7850-9}{{\em Eur. Phys. J. C} {\bfseries 80} no.~4, (2020) 347}, \href{http://arxiv.org/abs/1912.09335}{{\ttfamily arXiv:1912.09335 [hep-ph]}}.

\bibitem{Harrison:2020nrv}
{\bfseries LATTICE-HPQCD} Collaboration, J.~Harrison, C.~T.~H. Davies, and A.~Lytle, ``{$R(J/\psi)$ and $B_c^- \rightarrow J/\psi \ell^-\bar{\nu}_\ell$ Lepton Flavor Universality Violating Observables from Lattice QCD},'' \href{http://dx.doi.org/10.1103/PhysRevLett.125.222003}{{\em Phys. Rev. Lett.} {\bfseries 125} no.~22, (2020) 222003}, \href{http://arxiv.org/abs/2007.06956}{{\ttfamily arXiv:2007.06956 [hep-lat]}}.

\bibitem{Hu:2019qcn}
X.-Q. Hu, S.-P. Jin, and Z.-J. Xiao, ``{Semileptonic decays $B_c \to (\eta_c,J/\psi) l \bar{\nu}_l $ in the ''PQCD + Lattice'' approach},'' \href{http://dx.doi.org/10.1088/1674-1137/44/2/023104}{{\em Chin. Phys. C} {\bfseries 44} no.~2, (2020) 023104}, \href{http://arxiv.org/abs/1904.07530}{{\ttfamily arXiv:1904.07530 [hep-ph]}}.

\end{thebibliography}\endgroup
\bibliographystyle{utphys}

\end{document}